\documentclass[aps,onecolumn,showpacs,groupedaddress]{revtex4-1}
\usepackage{amssymb}
\usepackage{mathrsfs}
\usepackage{pstricks}
\usepackage{color}
\usepackage{graphicx}
\usepackage[fleqn]{amsmath}
\usepackage{slashed}
\usepackage{amsmath}
\usepackage{mathtools}
\usepackage{array}
\usepackage{tabularx}
\usepackage{physics}
\usepackage{slashed}
\usepackage[utf8]{inputenc}
\usepackage[T1]{fontenc}
\usepackage{float}

\begin{document}


\title{\bf On the completeness of the $\delta_{KLS}$-generalized statistical field theory}


\author{P. R. S. Carvalho}
\email{prscarvalho@ufpi.edu.br}
\affiliation{\it Departamento de F\'\i sica, Universidade Federal do Piau\'\i, 64049-550, Teresina, PI, Brazil}





\begin{abstract}
In this work we introduce a field-theoretic tool that enable us to evaluate the critical exponents of $\delta_{KLS}$-generalized systems undergoing continuous phase transitions, namely $\delta_{KLS}$-generalized statistical field theory. It generalizes the standard Boltzmann-Gibbs through the introduction of the $\delta_{KLS}$ parameter from which Boltzmann-Gibbs statistics is recovered in the limit $\delta_{KLS}\rightarrow 0$. From the results for the critical exponents we provide the referred physical interpretation for the $\delta_{KLS}$ parameter. Although new generalized universality classes emerge, we show that they are incomplete for describing the behavior of some real materials. This task is fulfilled only for nonextensive statistical field theory, which is related to fractal derivative and multifractal geometries, up to the moment, for our knowledge.
\end{abstract}


\maketitle


\section{Introduction}

\par In the last years, some generalized statistics were proposed for describing the physical behavior of many physical systems not described by Boltzmann-Gibbs statistics. For a list of them see Ref. \cite{Tsallis2023}. The original motivation for proposing these generalized statistics came from fractal calculus \cite{FractalDerivatives} and studies on the probabilistic description of multifractal geometries \cite{TSALLIS1999}. In fact, probabilistic description of conventional geometries, leading to the definition of conventional Boltzmann-Gibbs statistics are not capable of explaining many experimental results. Only by considering the probabilistic description of multifractal geometries and the corresponding definition of some generalized statistics is suffice to describe such experimental results \cite{TSALLIS1998534}. In the road for finding only the consistent ones, we have already shown that the corresponding field theories for a few of them are not capable of explaining the values of measured critical exponents for some real materials \cite{ALVES2023138005,CARVALHO2023138187}. On the other hand, the only generalized field theory that has attained such a task was nonextensive statistical field theory (NSFT), up to the moment, for our knowledge \cite{CARVALHO2023137683}. In fact, NSFT was successful in obtaining the critical exponents for some real, non-ideal materials presenting impurities, defects, inhomogeneities and competition among these effects \cite{Magnetochemistry.Turki,KHELIFI2014149,PhysRevB.75.024419,OMRI20123122,Ghosh_2005,doi:10.1063/1.2795796,GHODHBANE2013558,J.Appl.Phys.A.Berger,PhysRevB.68.144408,BenJemaa,PhysRevB.70.104417,PhysRevB.79.214426,J.Appl.Phys.Vasiliu-Doloca,YU2018393,PhysRevB.92.024409,ZHANG2013146,PHAN201440,RSCAdvJeddi,HCINI20152042,Phys.SolidStateBaazaoui}. It is the generalization of the Boltzmann-Gibbs based field-theoretic renormalization group introduced by Kenneth Wilson \cite{PhysRevLett.28.240} for describing the critical behavior of ideal, homogeneous, pure and perfect systems undergoing continuous phase transitions. As consistence constraints to be satisfied for a given field theory, we have that it has to emerge from a maximum principle, a trace-form entropy, present decisivity, maximality, concavity, Lesche stability, positivity, continuity, symmetry and expansibility \cite{PhysRevE.71.046128,Eur.Phys.J.B70.3}. The generalized statistics approached in the present work, namely $\delta_{KLS}$-statistical field theory ($\delta_{KLS}$-SFT), satisfies all the requirements aforementioned where $-1/3 < \delta_{KLS} < 1/3$ \cite{PhysRevE.71.046128}. We take the freedom for using $\delta_{KLS}$ to represent such a distribution once in the referred Ref., namely in Ref. \cite{PhysRevE.71.046128}, no symbol was attributed to it. Moreover, a generalized statistics has to describe all results for measured physical quantities in real materials. If the corresponding generalized statistics does not attain that goal for at least one material, then it is not so general and must be discarded as one trying to generalize statistical mechanics.  

\par In this work we build a generalized version of by Kenneth Wilson \cite{PhysRevLett.28.240} field-theoretic renormalization group, namely  statistical field theory (SFT). We expect to recover the Wilson's results for the critical exponents in the limit $\delta_{KLS}\rightarrow 0$. We also expect that there is a $\delta_{KLS}$-generalized universality class, namely the O($N$)$_{\delta_{KLS}}$ one. As particular cases, we have the $\delta_{KLS}$-generalized Ising and $\delta_{KLS}$-generalized Heisenberg universality classes. In these generalized universality classes, the $\delta_{KLS}$-generalized critical exponents depend on the dimension $d$, $N$ and symmetry of some $N$-component order parameter, if the interactions of their degrees of freedom are of short- or long-range type and $\delta_{KLS}$. From the results for the critical exponents we will provide the referred physical interpretation for the $\delta_{KLS}$ parameter. 

\par In this work we introduce the generalized $\delta_{KLS}$-SFT in Sect. \ref{deltaKLS}. In Sect. \ref{Somemodels} we display the results for the critical exponents for some models. We compare the theoretic and experimental results in Sect \ref{Comparison}. In Sect. \ref{Conclusions} we present our conclusions.

\section{$\delta_{KLS}$-SFT}\label{deltaKLS}

\par We introduce the $\delta_{KLS}$-SFT by defining its generating functional by
\begin{eqnarray}\label{huyhtrjisd}
Z[J] = \mathcal{N}^{-1}\exp_{\delta_{KLS}}\left[-\int d^{d}x\mathcal{L}_{int}\left(\frac{\delta}{\delta J(x)}\right)\right]\int\exp\left[\frac{1}{2}\int d^{d}xd^{d}x^{\prime}J(x)G_{0}(x-x^{\prime})J(x^{\prime})\right]
\end{eqnarray}
where we can determine the constant $\mathcal{N}$ from the condition $Z[J=0] = 1$,
\begin{eqnarray}
e_{\delta_{KLS}}^{x} = \left[A - \frac{1}{2}\sqrt{B + C + D} \right]^{1/\delta_{KLS}},
\end{eqnarray}
where $e_{\delta_{KLS}}^{x}$ is the $\delta_{KLS}$-exponential function \cite{PhysRevE.71.046128,Eur.Phys.J.B70.3} and
\begin{eqnarray}
A = -\frac{\sqrt{\frac{\sqrt[3]{\sqrt{3}\sqrt{27\delta_{KLS}^{4}x^{4} + 1} + 9\delta_{KLS}^{2}x^{2}}}{3^{2/3}} - \frac{1}{\sqrt[3]{3}\sqrt[3]{\sqrt{3}\sqrt{27\delta_{KLS}^{4}x^{4} + 1} + 9\delta_{KLS}^{2}x^{2}}}}}{\sqrt{2}},
\end{eqnarray}
\begin{eqnarray}
B = -\frac{4\sqrt{2}\delta_{KLS}x}{\sqrt{\frac{\sqrt[3]{\sqrt{3}\sqrt{27\delta_{KLS}^{4}x^{4} + 1} + 9\delta_{KLS}^{2}x^{2}}}{3^{2/3}} - \frac{1}{\sqrt[3]{3}\sqrt[3]{\sqrt{3}\sqrt{27\delta_{KLS}^{4}x^{4} + 1} + 9\delta_{KLS}^{2}x^{2}}}}},
\end{eqnarray}
\begin{eqnarray}
C = - \frac{2\sqrt[3]{\sqrt{3}\sqrt{27\delta_{KLS}^{4}x^{4} + 1} + 9\delta_{KLS}^{2}x^{2}}}{3^{2/3}},
\end{eqnarray}
\begin{eqnarray}
D = \frac{2}{\sqrt[3]{3}\sqrt[3]{\sqrt{3}\sqrt{27\delta_{KLS}^{4}x^{4} + 1} + 9\delta_{KLS}^{2}x^{2}}}.
\end{eqnarray}
The definition of Eq. (\ref{huyhtrjisd}) is inspired by the earlier work on NSFT \cite{CARVALHO2023137683}. The first exponential is $\delta_{KLS}$-generalized and is related to the interaction of the theory while the last one, corresponding to the free theory, is nongeneralized. This is analogous to NSFT. In fact, in NSFT, the nonextensive free theory depends on the nonextensive parameter $q$ in a nonlinear manner \cite{CARVALHO2023137683} thus not permitting us to define a free propagator as a linear superposition of creator and destruction operators. Then, if we can employ some approach based on a propagator, the free theory has to be extensive ($q = 1$) which is linear and permitting us to write the propagator as a linear superposition of creator and destruction operators. In the case of this work, the free theory represented by the last exponential of Eq. (\ref{huyhtrjisd}) has to be nongeneralized. Now we display the results for the critical exponents of some $\delta_{KLS}$-generalized models.

\section{Some models}\label{Somemodels}

\par Critical exponents without subscript are nongeneralized ones \cite{Wilson197475,Bonfirm_1981,PhysRevD.95.085001,PhysRevD.103.116024,STEPHEN197389,PhysRevE.60.2071,Hager_2002,BrezinEandParisiGandRicci-TersenghiF,LohmannMSladeGLallaceBC,SladeG,PhysRevLett.29.917,PhysRevD.10.3235,PhysRevD.94.125028,PhysRevB.13.251,PhysRev.86.821,PhysRevLett.28.240,PhysRevLett.35.1678,PhysRevB.67.104415,PhysRevB.72.224432,Albuquerque_2001,LEITE2004281,PhysRevB.61.14691,PhysRevB.68.052408,FARIAS,Borba,Santos_2014,deSena_2015,Santos_2019,Santos_20192,Leite_2022,PhysRevB.31.379,C.ItzyksonJ.M.Drouffe,ZINNJUSTIN1991105,PhysRevD.96.096010,JANSSEN2005147,Tauber_2005} and valid for all loop orders.

\subsection{Ising-like models}

\par
\begin{figure}[h]
\centering
\includegraphics[width=0.6\textwidth]{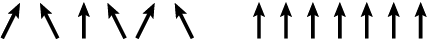}
\caption{Transition from a disordered phase (left) to some ordered one (right).}\label{Ising}
\end{figure}

 O($N$)$_{\delta_{KLS}}$ $\delta_{KLS}$-$\phi^{4}$ theory in $d = 4 - \epsilon$ dimensions:
\begin{eqnarray}\label{etaphi4}
\eta_{\delta_{KLS}} = \eta + \frac{2\delta_{KLS}}{1 - 2\delta_{KLS}}\frac{(N + 2)\epsilon^{2}}{2(N + 8)^{2}},  
\end{eqnarray}
\begin{eqnarray}\label{nuphi4}
\nu_{\delta_{KLS}} = \nu + \frac{2\delta_{KLS}}{1 - 2\delta_{KLS}}\frac{(N + 2)\epsilon}{4(N + 8)}
\end{eqnarray}
\begin{eqnarray}\label{z}
z_{\delta_{KLS}} = z + \frac{2\delta_{KLS}}{1 - 2\delta_{KLS}}\frac{[6\ln(4/3) - 1](N + 2)}{2(N + 8)^{2}}  \epsilon^{2},
\end{eqnarray}
where $N$ is the number of components of the field $\phi$. Fig. \ref{Ising} represents the transition of some Ising ($N = 1$) spin system from disordered (left) to ordered (right) phase.

\subsection{Both $\delta_{KLS}$-percolation and $\delta_{KLS}$-Yang-lee edge singularity}

\par Both $\delta_{KLS}$-percolation \cite{Bonfirm_1981} ($\alpha = -1$ and $\beta = -2$) and $\delta_{KLS}$-Yang-Lee edge singularity \cite{Bonfirm_1981} ($\alpha = -1$ and $\beta = -1$) in dimensions $d = 6 - \epsilon$: 
\begin{eqnarray}
\eta_{\delta_{KLS}} = \eta - \frac{\delta_{KLS}(5\delta_{KLS} - 4)}{1 - 2\delta_{KLS}} \frac{4\alpha\beta}{3(\alpha - 4\beta)[\alpha - 4\beta(1 - \delta_{KLS})(1 - 5\delta_{KLS})/(1 - 2\delta_{KLS})]}\epsilon ,
\end{eqnarray}
\begin{eqnarray}
\nu_{\delta_{KLS}}^{-1} = \nu^{-1} - \frac{\delta_{KLS}(5\delta_{KLS} - 4)}{1 - 2\delta_{KLS}} \frac{20\alpha\beta}{3(\alpha - 4\beta)[\alpha - 4\beta(1 - \delta_{KLS})(1 - 5\delta_{KLS})/(1 - 2\delta_{KLS})]}\epsilon ,
\end{eqnarray}
\begin{eqnarray}
\omega_{\delta_{KLS}} = \omega .
\end{eqnarray}
\begin{figure}[h]
\centering
\includegraphics[width=0.2\textwidth]{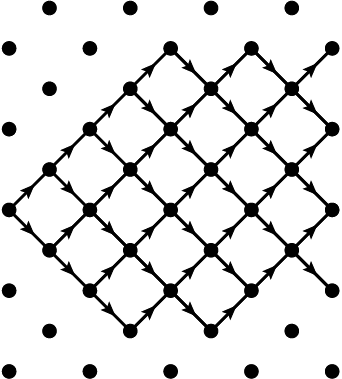}
\caption{Percolation phenomenon.}\label{percolation}
\end{figure}
Fig. \ref{percolation} represents the percolation phenomenon.

\subsection{$\delta_{KLS}$-$\phi^{6}$ theory}

\par $N$-component $\delta_{KLS}$-$\phi^{6}$ theory in $d = 3 - \epsilon$ dimensions: 
\begin{eqnarray}
\eta_{\delta_{KLS}} = \eta + \frac{2\delta_{KLS}}{1 - 2\delta_{KLS}}\frac{(N + 2)(N + 4)}{12(3N + 22)^{2}}\epsilon^{2}, 
\end{eqnarray}
\begin{eqnarray}
\nu_{\delta_{KLS}} = \nu + \frac{2\delta_{KLS}}{1 - 2\delta_{KLS}}\frac{(N + 2)(N + 4)}{3(3N + 22)^{2}}\epsilon^{2}.
\end{eqnarray}
where $N$ is the number of components of the field $\phi$.

\subsection{$\delta_{KLS}$-long-range systems}

\par The $\delta_{KLS}$-long-range $N$-component $\delta_{KLS}$-Ising-like models with interactions decaying as $1/r_{ij}^{d + \sigma}$ in $d = 2\sigma - \varepsilon$ dimensions:
\[
\eta_{\sigma, \hspace{.5mm}\delta_{KLS}} = 
     \begin{cases}
       \eta_{\delta_{KLS}} &\quad\text{if}\quad \sigma > 2 - \eta_{\delta_{KLS}} \\
       \eta_{\sigma}  &\quad\text{if}\quad d/2 < \sigma < 2 - \eta_{\delta_{KLS}} \\
       0 &\quad\text{if}\quad \sigma < d/2 , \\
     \end{cases}
\]
\[
\nu_{\sigma, \hspace{.5mm}\delta_{KLS}} = 
     \begin{cases}
       \nu_{\delta_{KLS}} &\quad\text{if}\quad \sigma > 2 - \eta_{\delta_{KLS}} \\
       \nu_{\sigma} + \frac{2\delta_{KLS}}{1 - 2\delta_{KLS}}\frac{(N + 2)}{\sigma^{2}(N + 8)}\epsilon &\quad\text{if}\quad d/2 < \sigma < 2 - \eta_{\delta_{KLS}} \\
       1/2 &\quad\text{if}\quad \sigma < d/2 , \\
     \end{cases}
\]
where $N$ is the number of components of the field $\phi$.

\subsection{$\delta_{KLS}$-Gross-Neveu model}

\par $\delta_{KLS}$-Gross-Neveu model in $d = 2 + \epsilon$ dimensions:
\begin{eqnarray}
\eta_{\psi, \hspace{.5mm}\delta_{KLS}} = \eta_{\psi} + \frac{2\delta_{KLS}}{1 - 2\delta_{KLS}}\frac{(2N - 1)}{8(N - 1)^{2}}\epsilon^{2}, 
\end{eqnarray}
\begin{eqnarray}
\eta_{\mathcal{M}, \hspace{.5mm}\delta_{KLS}} = \eta_{\mathcal{M}} + \frac{2\delta_{KLS}}{1 - 2\delta_{KLS}}\frac{\epsilon}{2(N - 1)}, 
\end{eqnarray}
\begin{eqnarray}
\nu_{\delta_{KLS}} = \nu ,
\end{eqnarray}
where $N$ is the number of components of the field $\phi$.

\subsection{$\delta_{KLS}$-uniaxial systems with strong dipolar forces}

\par $\delta_{KLS}$-uniaxial systems with strong dipolar forces in the $z$-direction in $d = 3 - \epsilon$ dimensions:
\begin{eqnarray}
\eta_{\delta_{KLS}} = \eta + \frac{2\delta_{KLS}}{1 - 2\delta_{KLS}}\frac{4(N + 2)}{9(N + 8)^{2}}\epsilon^{2},  
\end{eqnarray}
\begin{eqnarray}
\nu_{\delta_{KLS}}^{-1} = \nu^{-1} - \frac{2\delta_{KLS}}{1 - 2\delta_{KLS}}\frac{(N + 2)}{(N + 8)}\epsilon ,
\end{eqnarray}
where $N$ is the number of components of the field $\phi$.

\subsection{$\delta_{KLS}$-spherical model}

\par The $\delta_{KLS}$-spherical model \cite{PhysRev.86.821} can be obtained by taking the limit $N\rightarrow\infty$ \cite{PhysRev.176.718} of the O($N$)$_{\delta_{KLS}}$ model of the present work. After taking this limit, we obtain in $d = 4 - \epsilon$ 
\begin{eqnarray}
\eta_{\delta_{KLS}} = \eta ,  
\end{eqnarray}
\begin{eqnarray}
\nu_{\delta_{KLS}} = \nu + \frac{\delta_{KLS}}{1 - \delta_{KLS}}\frac{\epsilon}{4},
\end{eqnarray}
where the corresponding nongeneralized values are exact, namely $\eta = 0$ and $\nu = 1/(2 - \epsilon)$ \cite{PhysRevLett.28.240}.

\subsection{$\delta_{KLS}$-Lifshitz critical points}

\par Higher character $\delta_{KLS}$-Lifshitz anisotropic, computed in the orthogonal approximation, and isotropic (computed in both orthogonal approximation and exactly):
\begin{eqnarray}
\eta_{n, \hspace{.5mm}\delta_{KLS}} = \eta_{n} + \frac{2\delta_{KLS}}{1 - 2\delta_{KLS}}n\frac{(N + 2)}{2(N + 8)^{2}}\epsilon_{n}^{2}, 
\end{eqnarray}
\begin{eqnarray}
\nu_{n, \hspace{.5mm}\delta_{KLS}} = \nu_{n} + \frac{2\delta_{KLS}}{1 - 2\delta_{KLS}}\frac{(N + 2)}{4n(N + 8)}\epsilon_{n},
\end{eqnarray}
where $\epsilon_{L} = 4 + \sum_{n = 2}^{L}[(n - 1)/n]m_{n} - d$, 
\begin{eqnarray}
\eta_{n, \hspace{.5mm}\delta_{KLS}} = \eta_{n} + \frac{2\delta_{KLS}}{1 - 2\delta_{KLS}}\frac{(N + 2)}{2n(N + 8)^{2}}\epsilon_{n}^{2}, 
\end{eqnarray}
\begin{eqnarray}
\nu_{n, \hspace{.5mm}\delta_{KLS}} = \nu_{n} + \frac{2\delta_{KLS}}{1 - 2\delta_{KLS}}\frac{(N + 2)}{4n^{2}(N + 8)}\epsilon_{n},
\end{eqnarray}
where $\epsilon_{L} = 4n - d$, 
\begin{eqnarray}
\eta_{n, \hspace{.5mm}\delta_{KLS}} = \eta_{n} + \frac{2\delta_{KLS}}{1 - 2\delta_{KLS}}\frac{(-1)^{n + 1}\Gamma(2n)^{2}(N + 2)}{\Gamma(n + 1)\Gamma(3n)(N + 8)^{2}}\epsilon_{n}^{2}, 
\end{eqnarray}
\begin{eqnarray}
\nu_{n, \hspace{.5mm}\delta_{KLS}} = \nu_{n} + \frac{2\delta_{KLS}}{1 - 2\delta_{KLS}}\frac{(N + 2)}{4n^{2}(N + 8)}\epsilon_{n},
\end{eqnarray}
where $\epsilon_{L} = 4n - d$ and $N$ is the number of components of the field $\phi$.
\begin{figure}[h]
\centering
\includegraphics[width=0.9\textwidth]{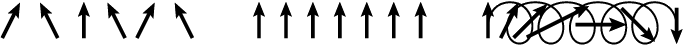}
\caption{Disordered (left), uniformly ordered (center) and modulated phases (right).}\label{Lifshitz}
\end{figure}
Fig. \ref{Lifshitz} represents the phases coexisting at a Lifshitz point.

\subsection{Long-range $\delta_{KLS}$-$\lambda\phi^{3}$ theory}

\par Long-range $\delta_{KLS}$-$\lambda\phi^{3}$ models with interactions decaying as $1/r_{ij}^{d + \sigma}$ in $d = 3\sigma - \varepsilon$ dimensions: 
\begin{eqnarray}
\eta_{\sigma , \hspace{.5mm}\delta_{KLS}} = \eta_{\sigma}, 
\end{eqnarray}
\begin{eqnarray}
\nu_{\sigma , \hspace{.5mm}\delta_{KLS}}^{-1} = \nu_{\sigma}^{-1} + \frac{\delta_{KLS}(5\delta_{KLS} - 4)}{(\delta_{KLS} - 1)(5\delta_{KLS} - 1)}\frac{\alpha}{2\beta}\epsilon .
\end{eqnarray}

\subsection{$\delta_{KLS}$-multicritical points}

\par $\delta_{KLS}$-multicritical points of order $k$, whose interactions are of the form $\phi^{2k}/(2k)!$ decaying as $1/r_{ij}^{d + \sigma}$, in dimensions $d = \frac{2k}{k - 1} - \epsilon$:
\begin{eqnarray}
\eta_{\delta_{KLS}} = \eta + \frac{2\delta_{KLS}}{1 - 2\delta_{KLS}}4(k - 1)^{2}\frac{(k)!^{6}}{(2k!)^{3}}\epsilon^{2}.
\end{eqnarray}
The situation in which we have multicritical points of order $k$, corresponds to a system in which $k$ phases can coexist at the same point, namely a multicritical point.

\subsection{$\delta_{KLS}$-Gross-Neveu-Yukawa model}

\par $\delta_{KLS}$-Gross-Neveu-Yukawa model in $d = 4 - \epsilon$ dimensions:   
\begin{eqnarray}
\eta_{\psi ,\hspace{.5mm}\delta_{KLS}} = \eta_{\psi} - \frac{1}{2}\delta_{KLS}(5\delta_{KLS} - 4) \frac{\epsilon}{(2N + 3)[(2N + 1)(1 - 2\delta_{KLS}) + 2(1 - \delta_{KLS})(1 - 5\delta_{KLS})]} ,
\end{eqnarray}
\begin{eqnarray}
\eta_{\phi ,\hspace{.5mm}\delta_{KLS}} = \eta_{\phi} - \delta_{KLS}(5\delta_{KLS} - 4) \frac{2\epsilon}{(2N + 3)[(2N + 1)(1 - 2\delta_{KLS}) + 2(1 - \delta_{KLS})(1 - 5\delta_{KLS})]} ,
\end{eqnarray}
\begin{eqnarray}
\nu_{\delta_{KLS}}^{-1} = \nu^{-1} - \frac{1}{(2N + 3)}\frac{A_{N,\hspace{.5mm}\delta_{KLS}}}{[(2N + 1)(1 - 2\delta_{KLS}) + 2(1 - \delta_{KLS})(1 - 5\delta_{KLS})]}\epsilon ,
\end{eqnarray}
\begin{eqnarray}
&& A_{N,\hspace{.5mm}\delta_{KLS}} = (2N + 3)[R_{N,\hspace{.5mm}\delta_{KLS}}/6(1 - 2\delta_{KLS}) + 2N(1 - 2\delta_{KLS})] - \nonumber \\&&   [(2N + 1)(1 - 2\delta_{KLS}) + 2(1 - \delta_{KLS})(1 - 5\delta_{KLS})](R_{N}/6 + 2N),
\end{eqnarray}
\begin{eqnarray}
&& R_{N,\hspace{.5mm}\delta_{KLS}} = - [(2N - 1)(1 - 2\delta_{KLS}) - 2(1 - \delta_{KLS})(1 - 5\delta_{KLS})] + \nonumber \\&&  \left\{[(2N - 1)(1 - 2\delta_{KLS}) - 2(1 - \delta_{KLS})(1 - 5\delta_{KLS})]^{2}  + \frac{144N}{1 - 2\delta_{KLS}}\right\}^{1/2},
\end{eqnarray}
where $N$ is the number of components of the field $\phi$.

\subsection{$\delta_{KLS}$-short- and long-range directed percolation}

\par Both $\delta_{KLS}$-short- and $\delta_{KLS}$-long-range directed percolation in $d = 4 - \epsilon$ and $d = 2\sigma - \varepsilon$ dimensions, respectively:  
\begin{eqnarray}\label{a}
\eta_{\delta_{KLS}} = \eta + \frac{\delta_{KLS}(5\delta_{KLS} - 4)}{(\delta_{KLS} - 1)(5\delta_{KLS} - 1)}\frac{\epsilon}{6},  
\end{eqnarray}
\begin{eqnarray}
\nu_{\delta_{KLS}} = \nu - \frac{\delta_{KLS}(5\delta_{KLS} - 4)}{(\delta_{KLS} - 1)(5\delta_{KLS} - 1)}\frac{\epsilon}{16}, 
\end{eqnarray}
\begin{eqnarray}\label{b}
z_{\delta_{KLS}} = z + \frac{\delta_{KLS}(5\delta_{KLS} - 4)}{(\delta_{KLS} - 1)(5\delta_{KLS} - 1)}\frac{\epsilon}{12}, 
\end{eqnarray}
\begin{eqnarray}\label{c}
\eta_{\sigma ,\hspace{.5mm}\delta_{KLS}} = \eta_{\sigma} + \frac{\delta_{KLS}(5\delta_{KLS} - 4)}{(\delta_{KLS} - 1)(5\delta_{KLS} - 1)}\frac{\varepsilon}{7},  
\end{eqnarray}
\begin{eqnarray}
\nu_{\sigma ,\hspace{.5mm}\delta_{KLS}} = \nu_{\sigma} - \frac{\delta_{KLS}(5\delta_{KLS} - 4)}{(\delta_{KLS} - 1)(5\delta_{KLS} - 1)}\frac{2\varepsilon}{7\sigma^{2}},
\end{eqnarray}
\begin{eqnarray}\label{d}
z_{\sigma ,\hspace{.5mm}\delta_{KLS}} = z_{\sigma} + \frac{\delta_{KLS}(5\delta_{KLS} - 4)}{(\delta_{KLS} - 1)(5\delta_{KLS} - 1)}\frac{\varepsilon}{7}.
\end{eqnarray}
\begin{figure}[h]
\centering
\includegraphics[width=0.2\textwidth]{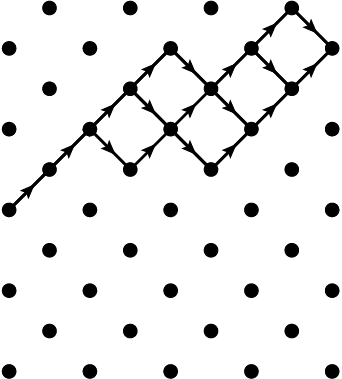}
\caption{Directed percolation phenomenon.}\label{directedpercolation}
\end{figure}
Fig. \ref{directedpercolation} represents the directed percolation phenomenon.

\subsection{$\delta_{KLS}$-short- and long-range dynamic isotropic percolation}

\par Both $\delta_{KLS}$-short- and $\delta_{KLS}$-long-range dynamic isotropic percolation at $d = 6 - \epsilon$ and $d = 3\sigma - \varepsilon$ dimensions, respectively: 
\begin{eqnarray}\label{e}
\eta_{\delta_{KLS}} = \eta + \frac{\delta_{KLS}(5\delta_{KLS} - 4)}{(\delta_{KLS} - 1)(5\delta_{KLS} - 1)}\frac{\epsilon}{21}, 
\end{eqnarray}
\begin{eqnarray}
\nu_{\delta_{KLS}} = \nu - \frac{\delta_{KLS}(5\delta_{KLS} - 4)}{(\delta_{KLS} - 1)(5\delta_{KLS} - 1)}\frac{5\epsilon}{84}, 
\end{eqnarray}
\begin{eqnarray}\label{f}
z_{\delta_{KLS}} = z + \frac{\delta_{KLS}(5\delta_{KLS} - 4)}{(\delta_{KLS} - 1)(5\delta_{KLS} - 1)}\frac{\epsilon}{6}, 
\end{eqnarray}
\begin{eqnarray}\label{g}
\eta_{\sigma ,\hspace{.5mm}\delta_{KLS}} = \eta_{\sigma} + \frac{\delta_{KLS}(5\delta_{KLS} - 4)}{(\delta_{KLS} - 1)(5\delta_{KLS} - 1)}\frac{3\varepsilon}{8}, 
\end{eqnarray}
\begin{eqnarray}
\nu_{\sigma ,\hspace{.5mm}\delta_{KLS}} = \nu_{\sigma} - \frac{\delta_{KLS}(5\delta_{KLS} - 4)}{(\delta_{KLS} - 1)(5\delta_{KLS} - 1)}\frac{\varepsilon}{4\sigma^{2}},
\end{eqnarray}
\begin{eqnarray}\label{h}
z_{\sigma ,\hspace{.5mm}\delta_{KLS}} = z_{\sigma} + \frac{\delta_{KLS}(5\delta_{KLS} - 4)}{(\delta_{KLS} - 1)(5\delta_{KLS} - 1)}\frac{3\varepsilon}{16}.
\end{eqnarray}
Eqs. (\ref{a}-\ref{b}), (\ref{c}-\ref{d}), (\ref{e}-\ref{f}), (\ref{g}-\ref{h}) represent the critical exponents for short-range directed percolation, long-range directed percolation, short-range dynamic isotropic percolation and long-range dynamic isotropic percolation, respectively.
\begin{figure}[h]
\centering
\includegraphics[width=0.2\textwidth]{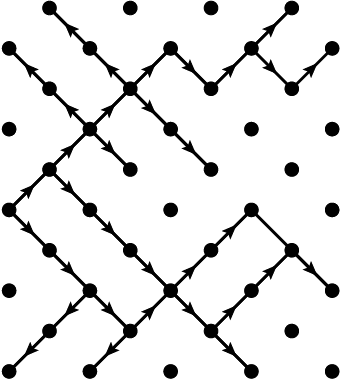}
\caption{Isotropic percolation phenomenon.}\label{isotropicpercolation}
\end{figure}
Fig. \ref{isotropicpercolation} represents the isotropic percolation phenomenon.

\section{Comparison between theoretic and experimental results}\label{Comparison}

\par Now we give the physical interpretation of the $\delta_{KLS}$ parameter. Consider some physical quantity near a continuous phase transition, for example the susceptibility. It diverges near the transition point. A measure of it divergence degree is given by the values of its critical exponent, namely $\gamma$. Its inverse gives an idea of how much the corresponding material will be susceptible to changes in the applied magnetic field. So higher (lower) values of $\gamma$ shows that how much more (less) susceptible or weaker (stronger) the constituents of the material interact among each other. In fact, we can obtain the effective energy of the system and obtain the same conclusions. By expanding around $\delta_{KLS} \approx 0$ and considering the leading term to the energy of the system, in units of $k_{B}T$, we have $e_{\delta_{KLS}}^{-E} \approx e^{-E}(1 - \delta_{KLS} E^{2}) \approx e^{-\left(E + \delta_{KLS} E^{2}\right)}$. Then the effective energy of the system is given by $E + \delta_{KLS} E^{2}$. It increases when we increase the values of $\delta_{KLS}$. So higher (lower) numerical values of $\delta_{KLS}$ indicate materials whose constituents interact weaker (stronger) or, equivalently, are more (less) susceptible. Thus, they have higher (lower) values for their critical indices. 

\par For evaluating numerical results from Eqs. (\ref{etaphi4})-(\ref{nuphi4}), we apply the following numerical values $\eta = 0.061(8)$, $\nu = 0.689(2)$, $\beta = 0.365(3)$, $\gamma = 1.336(4)$ \cite{PHAN2010238} and $\eta = 0.030(4)$, $\nu = 0.630(1)$, $\beta = 0.325(2)$, $\gamma = 1.241(2)$ \cite{PHAN2010238} obtained from experiments for nearly ideal crystals for Heinsenberg and Ising systems, respectively. The corresponding results obtained from experiments for some non-ideal manganites are in Tables \ref{tableexponentsE3N1}-\ref{tableexponentsE3N3}. In the case of $\delta_{KLS}$-Heisenberg systems (Table \ref{tableexponentsE3N3}), we obtain $ 0.339(3) < \beta_{\delta_{KLS}} < 0.505(4)$ and $ 1.253(6) < \gamma_{\delta_{KLS}} < 1.739(8)$.  Although the $\delta_{KLS}$-SFT can explain the most of the results shown in the Table \ref{tableexponentsE3N3}, it does not explain the results for the following materials: La$_{0.67}$Ca$_{0.33}$Mn$_{0.95}$Fe$_{0.05}$O$_{3}$\cite{NISHA201266}, La$_{0.67}$Ca$_{0.33}$Mn$_{0.90}$Cr$_{0.10}$O$_{3}$\cite{NISHA201240} and La$_{0.67}$Ca$_{0.33}$Mn$_{0.75}$Cr$_{0.25}$O$_{3}$\cite{NISHA201240} of that Table. Then $\delta_{KLS}$-\text{SFT} is not complete and must be discarded as one trying to generalize SFT. Only NSFT of Ref. \cite{CARVALHO2023137683}, up to the moment to our knowledge, remains as a fully consistent or complete generalized formulation of SFT \cite{PhysRevLett.28.240}. 

\par
\begin{figure}[h]
\centering
\includegraphics[width=.5\textwidth,]{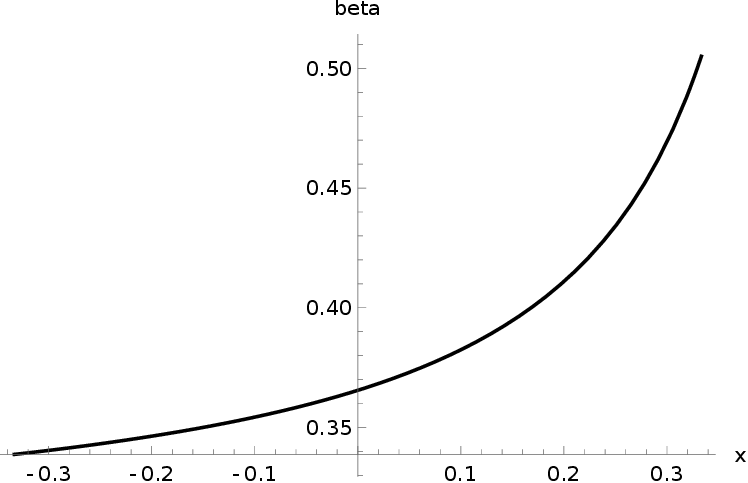}
\caption{Plot of the values for the critical exponent $\beta_{\delta_{KLS}}$ as a function of $\delta_{KLS}$ ($x$ axis).}\label{beta}
\end{figure}

\begin{figure}[h]
\centering
\includegraphics[width=.5\textwidth,]{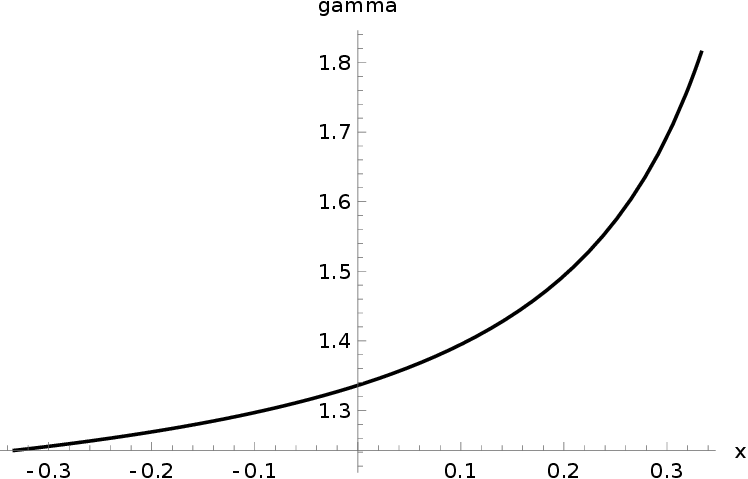}
\caption{Plot of the values for the critical exponent $\gamma_{\delta_{KLS}}$ as a function of $\delta_{KLS}$ ($x$ axis).}\label{gamma}
\end{figure}

 For all the systems treated in this work, we see the same behavior: they give a measure of how much the corresponding systems diverge near the transition point. Then, higher (lower) values of the critical exponents imply some weaker (stronger) effective interaction among their constituents. A plot of the results of Table \ref{tableexponentsE3N3} (less than those for the materials La$_{0.67}$Ca$_{0.33}$Mn$_{0.95}$Fe$_{0.05}$O$_{3}$\cite{NISHA201266}, La$_{0.67}$Ca$_{0.33}$Mn$_{0.90}$Cr$_{0.10}$O$_{3}$\cite{NISHA201240} and La$_{0.67}$Ca$_{0.33}$Mn$_{0.75}$Cr$_{0.25}$O$_{3}$\cite{NISHA201240} of that Table) is given in Figures (\ref{beta})-(\ref{gamma}).

\par As another application of the problem of this work, we mention one that considers correlations in chaos-coherence within chaotic phenomena of fluid at finite temperature \cite{BARY2022112572}. Similarly to the case of NSFT \cite{CARVALHO2023137683}, where the $q$-parameter can be interpreted as one encoding global correlations among the degrees of freedom of the system, the $\delta_{KLS}$-parameter also is related to correlations. Then, as the problem approached in Ref. \cite{BARY2022112572} investigates correlations, these correlations can be associated to the the $\delta_{KLS}$-parameter (although this task is not within the scope of the present work).

\begin{table}[H]
\caption{Static $\delta_{KLS}$-generalized critical exponents ($\delta_{KLS} \neq 0$) to $3$d ($\epsilon = 1$) Ising ($N = 1$) systems, obtained from experiment through Modified Arrott (MA) plots \cite{PhysRevLett.19.786}, Kouvel-Fisher (KF) method \cite{PhysRev.136.A1626} and $\delta_{KLS}$-SFT of this work.}
\begin{tabular}{p{7.0cm}p{3.0cm}p{1.0cm}} 
 \hline
 $\delta_{KLS}$-Ising & $\beta_{\delta_{KLS}}$ & $\gamma_{\delta_{KLS}}$    \\
 \hline
 La$_{0.8}$Sr$_{0.2}$MnO$_{3}$\cite{J.Appl.Phys.Vasiliu-Doloca}  &   0.290(10) &        \\
 \hline
 Nd$_{0.55}$Sr$_{0.45}$Mn$_{0.98}$Ga$_{0.02}$O$_{3}$\cite{YU2018393}  & 0.308(10) &  1.197 \\
 \hline
 Pr$_{0.6}$Sr$_{0.4}$MnO$_{3}$\cite{PhysRevB.92.024409}MAP  &   0.315(0) &  1.095(7)      \\
 Pr$_{0.6}$Sr$_{0.4}$MnO$_{3}$\cite{PhysRevB.92.024409}KF  &   0.312(2) &  1.106(5)      \\
 \hline
 La$_{0.8}$Ca$_{0.2}$MnO$_{3}$\cite{ZHANG2013146}KF  &   0.316(7) &  1.081(36)      \\
 \hline
 La$_{0.7}$Ca$_{0.3}$Mn$_{0.85}$Ni$_{0.15}$O$_{3}$\cite{PHAN201440}MAP  &   0.320(9) &  0.990(82)      \\
 \hline
 Nd$_{0.6}$Sr$_{0.4}$MnO$_{3}$\cite{RSCAdvJeddi}MAP  &   0.320(6) &  1.239(2)      \\
 Nd$_{0.6}$Sr$_{0.4}$MnO$_{3}$\cite{RSCAdvJeddi}KF  &   0.323(2) &  1.235(4)      \\
 \hline
 Nd$_{0.6}$Sr$_{0.4}$MnO$_{3}$\cite{PhysRevB.92.024409}MAP  &   0.321(3) &  1.183(17)      \\
 Nd$_{0.6}$Sr$_{0.4}$MnO$_{3}$\cite{PhysRevB.92.024409}KF  &   0.308(4) &  1.172(11)      \\
 \hline
 Nd$_{0.67}$Ba$_{0.33}$MnO$_{3}$\cite{HCINI20152042}MAP  &   0.325(4) &  1.248(19)      \\
 Nd$_{0.67}$Ba$_{0.33}$MnO$_{3}$\cite{HCINI20152042}KF  &   0.326(5) &  1.244(33)      \\
 \hline
 La$_{0.65}$Bi$_{0.05}$Sr$_{0.3}$MnO$_{3}$\cite{Phys.SolidStateBaazaoui}MAP  &   0.335(3) &  1.207(20)      \\
 La$_{0.65}$Bi$_{0.05}$Sr$_{0.3}$MnO$_{3}$\cite{Phys.SolidStateBaazaoui}KF  &   0.316(7) &  1.164(20)      \\
 \hline
 La$_{0.65}$Bi$_{0.05}$Sr$_{0.3}$Mn$_{0.94}$Ga$_{0.06}$O$_{3}$\cite{Phys.SolidStateBaazaoui}MAP & 0.334(4) & 1.192(8) \\
 La$_{0.65}$Bi$_{0.05}$Sr$_{0.3}$Mn$_{0.94}$Ga$_{0.06}$O$_{3}$\cite{Phys.SolidStateBaazaoui}KF  & 0.307(8) & 1.138(5) \\
 \hline
\end{tabular}
\label{tableexponentsE3N1}
\end{table}

\begin{table}[H]
\caption{Static $\delta_{KLS}$-generalized critical exponents ($\delta_{KLS} \neq 0$) to $3$d ($\epsilon = 1$) Heisenberg ($N = 3$) systems, obtained from experiment through Modified Arrott (MA) plots \cite{PhysRevLett.19.786}, Kouvel-Fisher (KF) method \cite{PhysRev.136.A1626} and $\delta_{KLS}$-SFT of this work.}
\begin{tabular}{p{7.0cm}p{3.0cm}p{1.0cm}} 
 \hline
$\delta_{KLS}$-Heisenberg & $\beta_{\delta_{KLS}}$ & $\gamma_{\delta_{KLS}}$    \\
 \hline
 La$_{0.67}$Sr$_{0.33}$MnO$_{3}$\cite{MNEFGUI2014193}MAP  &   0.333(8) &  1.325(1)    \\
 \hline
 Pr$_{0.77}$Pb$_{0.23}$MnO$_{3}$\cite{PhysRevB.75.024419}MAP  &   0.343(5) &  1.357(20)     \\
 Pr$_{0.77}$Pb$_{0.23}$MnO$_{3}$\cite{PhysRevB.75.024419}KF  &   0.344(1) &  1.352(6)     \\
 \hline
 AMnO$_{3}$\cite{OMRI20123122}MAP  &   0.355(7) &  1.326(2)     \\
 AMnO$_{3}$\cite{OMRI20123122}KF  &   0.344(5) &  1.335(2)     \\
 \hline
 Nd$_{0.7}$Pb$_{0.3}$MnO$_{3}$\cite{Ghosh_2005}MAP  &   0.361(13) &  1.325(1)           \\
 Nd$_{0.7}$Pb$_{0.3}$MnO$_{3}$\cite{Ghosh_2005}KF  &   0.361(5) &  1.314(1)           \\
 \hline
 LaTi$_{0.2}$Mn$_{0.8}$O$_{3}$\cite{doi:10.1063/1.2795796}KF  &   0.359(4) &  1.280(10)     \\
 \hline
 La$_{0.67}$Sr$_{0.33}$Mn$_{0.95}$V$_{0.05}$O$_{3}$\cite{MNEFGUI2014193}MAP  &   0.358(5) &  1.381(4)  \\
 Nd$_{0.85}$Pb$_{0.15}$MnO$_{3}$\cite{Ghosh_2005}MAP  &  0.372(1) &  1.340(30)  \\
 Nd$_{0.85}$Pb$_{0.15}$MnO$_{3}$\cite{Ghosh_2005}KF  &  0.372(4) &  1.347(1)  \\
 \hline
 Nd$_{0.6}$Pb$_{0.4}$MnO$_{3}$\cite{PhysRevB.68.144408}KF  &   0.374(6) &  1.329(3)  \\
 \hline
 La$_{0.67}$Sr$_{0.33}$Mn$_{0.95}$V$_{0.15}$O$_{3}$\cite{MNEFGUI2014193}MAP  &   0.375(3) &  1.355(6)  \\
 \hline
 La$_{0.67}$Ba$_{0.22}$Sr$_{0.11}$MnO$_{3}$\cite{BenJemaa}MAP  &   0.378(3) &  1.388(1)   \\
 La$_{0.67}$Ba$_{0.22}$Sr$_{0.11}$MnO$_{3}$\cite{BenJemaa}KF  &   0.386(6) &  1.393(4)   \\
 \hline
 LaTi$_{0.95}$Mn$_{0.05}$O$_{3}$\cite{doi:10.1063/1.2795796}KF  &   0.378(7) &  1.290(20)     \\
 \hline
 LaTi$_{0.9}$Mn$_{0.1}$O$_{3}$\cite{doi:10.1063/1.2795796}KF  &   0.375(5) &  1.250(20)     \\
 \hline
 LaTi$_{0.85}$Mn$_{0.15}$O$_{3}$\cite{doi:10.1063/1.2795796}KF  &   0.376(3) &  1.240(10)     \\
 \hline
 La$_{0.67}$Ca$_{0.33}$Mn$_{0.9}$Ga$_{0.1}$O$_{3}$\cite{PhysRevB.70.104417}MAP  &  0.380(2) &  1.365(8)  \\
 La$_{0.67}$Ca$_{0.33}$Mn$_{0.9}$Ga$_{0.1}$O$_{3}$\cite{PhysRevB.70.104417}KF  &   0.387(6) &  1.362(2)  \\
 \hline
 La$_{0.67}$Ba$_{0.22}$Sr$_{0.11}$Mn$_{0.9}$Fe$_{0.1}$O$_{3}$\cite{BenJemaa}MAP  &   0.398(2) &  1.251(5)   \\
 La$_{0.67}$Ba$_{0.22}$Sr$_{0.11}$Mn$_{0.9}$Fe$_{0.1}$O$_{3}$\cite{BenJemaa}KF  &   0.395(3) &  1.247(3)   \\
 \hline
 La$_{0.67}$Ba$_{0.22}$Sr$_{0.11}$Mn$_{0.8}$Fe$_{0.2}$O$_{3}$\cite{BenJemaa}MAP  &   0.411(1) &  1.241(4)   \\
 La$_{0.67}$Ba$_{0.22}$Sr$_{0.11}$Mn$_{0.8}$Fe$_{0.2}$O$_{3}$\cite{BenJemaa}KF  &   0.394(3) &  1.292(3)   \\
 \hline
 La$_{0.7}$Sr$_{0.3}$Mn$_{0.99}$Ni$_{0.01}$O$_{3}$\cite{GINTING201317}KF  &   0.394(15) &  1.092(47)     \\
 \hline
 Pr$_{0.7}$Pb$_{0.3}$MnO$_{3}$\cite{PhysRevB.75.024419}MAP  &   0.404(6) &  1.354(20)     \\
 Pr$_{0.7}$Pb$_{0.3}$MnO$_{3}$\cite{PhysRevB.75.024419}KF  &   0.404(1) &  1.357(6)     \\
 \hline
 La$_{0.7}$Sr$_{0.3}$Mn$_{0.92}$Ni$_{0.02}$O$_{3}$\cite{GINTING201317}KF  &   0.400(17) &  1.081(32)     \\
 \hline
 La$_{0.75}$(Sr,Ca)$_{0.25}$Mn$_{0.9}$Ga$_{0.1}$O$_{3}$\cite{BenJemaa}MAP  &   0.420(5) &  1.221(2)   \\
 La$_{0.75}$(Sr,Ca)$_{0.25}$Mn$_{0.9}$Ga$_{0.1}$O$_{3}$\cite{BenJemaa}KF  &   0.428(5) &  1.286(4)   \\
 \hline
 Pr$_{0.5}$Sr$_{0.5}$MnO$_{3}$\cite{PhysRevB.79.214426}MAP  &   0.443(2) &  1.339(6)     \\
 Pr$_{0.5}$Sr$_{0.5}$MnO$_{3}$\cite{PhysRevB.79.214426}KF  &   0.448(9) &  1.334(10)     \\
 \hline
 La$_{0.7}$Sr$_{0.3}$Mn$_{0.97}$Ni$_{0.03}$O$_{3}$\cite{GINTING201317}KF  &   0.468(6) &  1.010(21)     \\
 \hline
 La$_{0.7}$Sr$_{0.3}$Mn$_{0.94}$Co$_{0.06}$O$_{3}$\cite{PHAN2014S247}MAP  &   0.478(13) &  1.165(27)     \\
 \hline
 La$_{0.7}$Sr$_{0.3}$Mn$_{0.92}$Co$_{0.08}$O$_{3}$\cite{PHAN2014S247}MAP  &   0.483(18) &  1.112(28)     \\
 \hline
 La$_{0.7}$Sr$_{0.3}$Mn$_{0.90}$Co$_{0.10}$O$_{3}$\cite{PHAN2014S247}MAP  &   0.487(16) &  1.109(63)     \\
 \hline
 La$_{0.67}$Ca$_{0.33}$Mn$_{0.95}$Fe$_{0.05}$O$_{3}$\cite{NISHA201266}MAP  &   0.550(10) &  1.0246(3)     \\
 \hline
 La$_{0.67}$Ca$_{0.33}$Mn$_{0.90}$Cr$_{0.10}$O$_{3}$\cite{NISHA201240}MAP  &   0.555(6) &   1.170(40)     \\
 \hline
 La$_{0.67}$Ca$_{0.33}$Mn$_{0.75}$Cr$_{0.25}$O$_{3}$\cite{NISHA201240}MAP  &  0.680(10) &  1.090(30) \\ 
 \hline
\end{tabular}
\label{tableexponentsE3N3}
\end{table}


\section{Conclusions}\label{Conclusions}

\par In this work we have introduced a field-theoretic tool, namely $\delta_{KLS}$-SFT for computing the critical exponents for non-ideal systems undergoing continuous phase transitions. We have showed that it is incomplete, \emph{i. e.}, it does not describe the behavior of some real non-ideal materials. Then it must be discarded as one trying to generalize SFT \cite{PhysRevLett.28.240}. This task remains being attained just for NSFT \cite{CARVALHO2023137683}, up to the moment to our knowledge, the only fully consistent or complete generalized formulation of SFT \cite{PhysRevLett.28.240}. For all the systems treated in this work, we see the same behavior: they give a measure of how much the corresponding systems diverge near the transition point. Then, higher (lower) values of the critical exponents imply some weaker (stronger) effective interaction among their constituents. A plot of the results of Table 2 (less than those for the materials La$_{0.67}$Ca$_{0.33}$Mn$_{0.95}$Fe$_{0.05}$O$_{3}$[68], La$_{0.67}$Ca$_{0.33}$Mn$_{0.90}$Cr$_{0.10}$O$_{3}$[68] and La$_{0.67}$Ca$_{0.33}$Mn$_{0.75}$Cr$_{0.25}$O$_{3}$[69] of that Table) is given in Figures (6)-(7).

\section*{Acknowledgments}

\par PRSC would like to thank the Brazilian funding agencies CAPES and CNPq (Grant: Produtividade 306130/2022-0) for ﬁnancial support.

\bibliography{apstemplate}

\providecommand{\noopsort}[1]{}\providecommand{\singleletter}[1]{#1}%
\begin{thebibliography}{75}%
\makeatletter
\providecommand \@ifxundefined [1]{%
 \@ifx{#1\undefined}
}%
\providecommand \@ifnum [1]{%
 \ifnum #1\expandafter \@firstoftwo
 \else \expandafter \@secondoftwo
 \fi
}%
\providecommand \@ifx [1]{%
 \ifx #1\expandafter \@firstoftwo
 \else \expandafter \@secondoftwo
 \fi
}%
\providecommand \natexlab [1]{#1}%
\providecommand \enquote  [1]{``#1''}%
\providecommand \bibnamefont  [1]{#1}%
\providecommand \bibfnamefont [1]{#1}%
\providecommand \citenamefont [1]{#1}%
\providecommand \href@noop [0]{\@secondoftwo}%
\providecommand \href [0]{\begingroup \@sanitize@url \@href}%
\providecommand \@href[1]{\@@startlink{#1}\@@href}%
\providecommand \@@href[1]{\endgroup#1\@@endlink}%
\providecommand \@sanitize@url [0]{\catcode `\\12\catcode `\$12\catcode
  `\&12\catcode `\#12\catcode `\^12\catcode `\_12\catcode `\%12\relax}%
\providecommand \@@startlink[1]{}%
\providecommand \@@endlink[0]{}%
\providecommand \url  [0]{\begingroup\@sanitize@url \@url }%
\providecommand \@url [1]{\endgroup\@href {#1}{\urlprefix }}%
\providecommand \urlprefix  [0]{URL }%
\providecommand \Eprint [0]{\href }%
\providecommand \doibase [0]{http://dx.doi.org/}%
\providecommand \selectlanguage [0]{\@gobble}%
\providecommand \bibinfo  [0]{\@secondoftwo}%
\providecommand \bibfield  [0]{\@secondoftwo}%
\providecommand \translation [1]{[#1]}%
\providecommand \BibitemOpen [0]{}%
\providecommand \bibitemStop [0]{}%
\providecommand \bibitemNoStop [0]{.\EOS\space}%
\providecommand \EOS [0]{\spacefactor3000\relax}%
\providecommand \BibitemShut  [1]{\csname bibitem#1\endcsname}%
\let\auto@bib@innerbib\@empty
\bibitem [{\citenamefont {Tsallis}(2023)}]{Tsallis2023}%
  \BibitemOpen
  \bibfield  {author} {\bibinfo {author} {\bibfnamefont {C.}~\bibnamefont
  {Tsallis}},\ }\href@noop {} {\emph {\bibinfo {title} {Introduction to
  Nonextensive Statistical Mechanics: Approaching a Complex World}}}\ (\bibinfo
   {publisher} {Springer, 2nd Edition},\ \bibinfo {year} {2023})\BibitemShut
  {NoStop}%
\bibitem [{\citenamefont {Deppman}\ \emph {et~al.}(2023)\citenamefont
  {Deppman}, \citenamefont {Megías},\ and\ \citenamefont
  {Pasechnik}}]{FractalDerivatives}%
  \BibitemOpen
  \bibfield  {author} {\bibinfo {author} {\bibfnamefont {A.}~\bibnamefont
  {Deppman}}, \bibinfo {author} {\bibfnamefont {E.}~\bibnamefont {Megías}}, \
  and\ \bibinfo {author} {\bibfnamefont {R.}~\bibnamefont {Pasechnik}},\
  }\href@noop {} {\bibfield  {journal} {\bibinfo  {journal} {Entropy}\ }\textbf
  {\bibinfo {volume} {25}},\ \bibinfo {pages} {1008} (\bibinfo {year}
  {2023})}\BibitemShut {NoStop}%
\bibitem [{\citenamefont {Tsallis}(1999)}]{TSALLIS1999}%
  \BibitemOpen
  \bibfield  {author} {\bibinfo {author} {\bibfnamefont {C.}~\bibnamefont
  {Tsallis}},\ }\href@noop {} {\bibfield  {journal} {\bibinfo  {journal}
  {{Braz. J. Phys.}}\ }\textbf {\bibinfo {volume} {29}},\ \bibinfo {pages} {1}
  (\bibinfo {year} {1999})}\BibitemShut {NoStop}%
\bibitem [{\citenamefont {Tsallis}\ \emph {et~al.}(1998)\citenamefont
  {Tsallis}, \citenamefont {Mendes},\ and\ \citenamefont
  {Plastino}}]{TSALLIS1998534}%
  \BibitemOpen
  \bibfield  {author} {\bibinfo {author} {\bibfnamefont {C.}~\bibnamefont
  {Tsallis}}, \bibinfo {author} {\bibfnamefont {R.}~\bibnamefont {Mendes}}, \
  and\ \bibinfo {author} {\bibfnamefont {A.}~\bibnamefont {Plastino}},\
  }\href@noop {} {\bibfield  {journal} {\bibinfo  {journal} {Physica A}\
  }\textbf {\bibinfo {volume} {261}},\ \bibinfo {pages} {534} (\bibinfo {year}
  {1998})}\BibitemShut {NoStop}%
\bibitem [{\citenamefont {Alves}\ \emph {et~al.}(2023)\citenamefont {Alves},
  \citenamefont {Neto}, \citenamefont {Lima}, \citenamefont {Alves},\ and\
  \citenamefont {Carvalho}}]{ALVES2023138005}%
  \BibitemOpen
  \bibfield  {author} {\bibinfo {author} {\bibfnamefont {T.~F.~A.}\
  \bibnamefont {Alves}}, \bibinfo {author} {\bibfnamefont {J.~F.~S.}\
  \bibnamefont {Neto}}, \bibinfo {author} {\bibfnamefont {F.~W.~S.}\
  \bibnamefont {Lima}}, \bibinfo {author} {\bibfnamefont {G.~A.}\ \bibnamefont
  {Alves}}, \ and\ \bibinfo {author} {\bibfnamefont {P.~R.~S.}\ \bibnamefont
  {Carvalho}},\ }\href@noop {} {\bibfield  {journal} {\bibinfo  {journal}
  {Phys. Lett. B}\ }\textbf {\bibinfo {volume} {843}},\ \bibinfo {pages}
  {138005} (\bibinfo {year} {2023})}\BibitemShut {NoStop}%
\bibitem [{\citenamefont {Carvalho}(2023{\natexlab{a}})}]{CARVALHO2023138187}%
  \BibitemOpen
  \bibfield  {author} {\bibinfo {author} {\bibfnamefont {P.~R.~S.}\
  \bibnamefont {Carvalho}},\ }\href@noop {} {\bibfield  {journal} {\bibinfo
  {journal} {Phys. Lett. B}\ }\textbf {\bibinfo {volume} {846}},\ \bibinfo
  {pages} {138187} (\bibinfo {year} {2023}{\natexlab{a}})}\BibitemShut
  {NoStop}%
\bibitem [{\citenamefont {Carvalho}(2023{\natexlab{b}})}]{CARVALHO2023137683}%
  \BibitemOpen
  \bibfield  {author} {\bibinfo {author} {\bibfnamefont {P.~R.~S.}\
  \bibnamefont {Carvalho}},\ }\href@noop {} {\bibfield  {journal} {\bibinfo
  {journal} {Phys. Lett. B}\ }\textbf {\bibinfo {volume} {838}},\ \bibinfo
  {pages} {137683} (\bibinfo {year} {2023}{\natexlab{b}})}\BibitemShut
  {NoStop}%
\bibitem [{\citenamefont {Turki}\ \emph {et~al.}(2017)\citenamefont {Turki},
  \citenamefont {Ghouri}, \citenamefont {Al-Meer}, \citenamefont {Elsaid},
  \citenamefont {Ahmad}, \citenamefont {Easa}, \citenamefont {Remenyi},
  \citenamefont {Mahmood}, \citenamefont {Hlil}, \citenamefont {Ellouze},\ and\
  \citenamefont {Elhalouani}}]{Magnetochemistry.Turki}%
  \BibitemOpen
  \bibfield  {author} {\bibinfo {author} {\bibfnamefont {D.}~\bibnamefont
  {Turki}}, \bibinfo {author} {\bibfnamefont {Z.~K.}\ \bibnamefont {Ghouri}},
  \bibinfo {author} {\bibfnamefont {S.}~\bibnamefont {Al-Meer}}, \bibinfo
  {author} {\bibfnamefont {K.}~\bibnamefont {Elsaid}}, \bibinfo {author}
  {\bibfnamefont {M.~I.}\ \bibnamefont {Ahmad}}, \bibinfo {author}
  {\bibfnamefont {A.}~\bibnamefont {Easa}}, \bibinfo {author} {\bibfnamefont
  {G.}~\bibnamefont {Remenyi}}, \bibinfo {author} {\bibfnamefont
  {S.}~\bibnamefont {Mahmood}}, \bibinfo {author} {\bibfnamefont {E.~K.}\
  \bibnamefont {Hlil}}, \bibinfo {author} {\bibfnamefont {M.}~\bibnamefont
  {Ellouze}}, \ and\ \bibinfo {author} {\bibfnamefont {F.}~\bibnamefont
  {Elhalouani}},\ }\href@noop {} {\bibfield  {journal} {\bibinfo  {journal}
  {Magnetochemistry}\ }\textbf {\bibinfo {volume} {3}},\ \bibinfo {pages} {28}
  (\bibinfo {year} {2017})}\BibitemShut {NoStop}%
\bibitem [{\citenamefont {Khelifi}\ \emph {et~al.}(2014)\citenamefont
  {Khelifi}, \citenamefont {Tozri}, \citenamefont {Dhahri},\ and\ \citenamefont
  {Hlil}}]{KHELIFI2014149}%
  \BibitemOpen
  \bibfield  {author} {\bibinfo {author} {\bibfnamefont {J.}~\bibnamefont
  {Khelifi}}, \bibinfo {author} {\bibfnamefont {A.}~\bibnamefont {Tozri}},
  \bibinfo {author} {\bibfnamefont {E.}~\bibnamefont {Dhahri}}, \ and\ \bibinfo
  {author} {\bibfnamefont {E.}~\bibnamefont {Hlil}},\ }\href@noop {} {\bibfield
   {journal} {\bibinfo  {journal} {J. Magn. Magn. Mater.}\ }\textbf {\bibinfo
  {volume} {349}},\ \bibinfo {pages} {149} (\bibinfo {year}
  {2014})}\BibitemShut {NoStop}%
\bibitem [{\citenamefont {Padmanabhan}\ \emph {et~al.}(2007)\citenamefont
  {Padmanabhan}, \citenamefont {Bhat}, \citenamefont {Elizabeth}, \citenamefont
  {R\"o\ss{}ler}, \citenamefont {R\"o\ss{}ler}, \citenamefont {D\"orr},\ and\
  \citenamefont {M\"uller}}]{PhysRevB.75.024419}%
  \BibitemOpen
  \bibfield  {author} {\bibinfo {author} {\bibfnamefont {B.}~\bibnamefont
  {Padmanabhan}}, \bibinfo {author} {\bibfnamefont {H.~L.}\ \bibnamefont
  {Bhat}}, \bibinfo {author} {\bibfnamefont {S.}~\bibnamefont {Elizabeth}},
  \bibinfo {author} {\bibfnamefont {S.}~\bibnamefont {R\"o\ss{}ler}}, \bibinfo
  {author} {\bibfnamefont {U.~K.}\ \bibnamefont {R\"o\ss{}ler}}, \bibinfo
  {author} {\bibfnamefont {K.}~\bibnamefont {D\"orr}}, \ and\ \bibinfo {author}
  {\bibfnamefont {K.~H.}\ \bibnamefont {M\"uller}},\ }\href@noop {} {\bibfield
  {journal} {\bibinfo  {journal} {Phys. Rev. B}\ }\textbf {\bibinfo {volume}
  {75}},\ \bibinfo {pages} {024419} (\bibinfo {year} {2007})}\BibitemShut
  {NoStop}%
\bibitem [{\citenamefont {Omri}\ \emph {et~al.}(2012)\citenamefont {Omri},
  \citenamefont {Tozri}, \citenamefont {Bejar}, \citenamefont {Dhahri},\ and\
  \citenamefont {Hlil}}]{OMRI20123122}%
  \BibitemOpen
  \bibfield  {author} {\bibinfo {author} {\bibfnamefont {A.}~\bibnamefont
  {Omri}}, \bibinfo {author} {\bibfnamefont {A.}~\bibnamefont {Tozri}},
  \bibinfo {author} {\bibfnamefont {M.}~\bibnamefont {Bejar}}, \bibinfo
  {author} {\bibfnamefont {E.}~\bibnamefont {Dhahri}}, \ and\ \bibinfo {author}
  {\bibfnamefont {E.}~\bibnamefont {Hlil}},\ }\href@noop {} {\bibfield
  {journal} {\bibinfo  {journal} {J. Magn. Magn. Mater.}\ }\textbf {\bibinfo
  {volume} {324}},\ \bibinfo {pages} {3122} (\bibinfo {year}
  {2012})}\BibitemShut {NoStop}%
\bibitem [{\citenamefont {Ghosh}\ \emph {et~al.}(2005)\citenamefont {Ghosh},
  \citenamefont {Rö{\ss}ler}, \citenamefont {Rö{\ss}ler}, \citenamefont
  {Nenkov}, \citenamefont {Elizabeth}, \citenamefont {Bhat}, \citenamefont
  {Dörr},\ and\ \citenamefont {Müller}}]{Ghosh_2005}%
  \BibitemOpen
  \bibfield  {author} {\bibinfo {author} {\bibfnamefont {N.}~\bibnamefont
  {Ghosh}}, \bibinfo {author} {\bibfnamefont {S.}~\bibnamefont {Rö{\ss}ler}},
  \bibinfo {author} {\bibfnamefont {U.~K.}\ \bibnamefont {Rö{\ss}ler}},
  \bibinfo {author} {\bibfnamefont {K.}~\bibnamefont {Nenkov}}, \bibinfo
  {author} {\bibfnamefont {S.}~\bibnamefont {Elizabeth}}, \bibinfo {author}
  {\bibfnamefont {H.~L.}\ \bibnamefont {Bhat}}, \bibinfo {author}
  {\bibfnamefont {K.}~\bibnamefont {Dörr}}, \ and\ \bibinfo {author}
  {\bibfnamefont {K.-H.}\ \bibnamefont {Müller}},\ }\href@noop {} {\bibfield
  {journal} {\bibinfo  {journal} {J. Phys. Condens. Matter}\ }\textbf {\bibinfo
  {volume} {18}},\ \bibinfo {pages} {557} (\bibinfo {year} {2005})}\BibitemShut
  {NoStop}%
\bibitem [{\citenamefont {Yang}\ and\ \citenamefont
  {Lee}(2007)}]{doi:10.1063/1.2795796}%
  \BibitemOpen
  \bibfield  {author} {\bibinfo {author} {\bibfnamefont {J.}~\bibnamefont
  {Yang}}\ and\ \bibinfo {author} {\bibfnamefont {Y.~P.}\ \bibnamefont {Lee}},\
  }\href@noop {} {\bibfield  {journal} {\bibinfo  {journal} {Appl. Phys.
  Lett.}\ }\textbf {\bibinfo {volume} {91}},\ \bibinfo {pages} {142512}
  (\bibinfo {year} {2007})}\BibitemShut {NoStop}%
\bibitem [{\citenamefont {Ghodhbane}\ \emph {et~al.}(2013)\citenamefont
  {Ghodhbane}, \citenamefont {Dhahri}, \citenamefont {Dhahri}, \citenamefont
  {Hlil}, \citenamefont {Dhahri}, \citenamefont {Alhabradi},\ and\
  \citenamefont {Zaidi}}]{GHODHBANE2013558}%
  \BibitemOpen
  \bibfield  {author} {\bibinfo {author} {\bibfnamefont {S.}~\bibnamefont
  {Ghodhbane}}, \bibinfo {author} {\bibfnamefont {A.}~\bibnamefont {Dhahri}},
  \bibinfo {author} {\bibfnamefont {N.}~\bibnamefont {Dhahri}}, \bibinfo
  {author} {\bibfnamefont {E.}~\bibnamefont {Hlil}}, \bibinfo {author}
  {\bibfnamefont {J.}~\bibnamefont {Dhahri}}, \bibinfo {author} {\bibfnamefont
  {M.}~\bibnamefont {Alhabradi}}, \ and\ \bibinfo {author} {\bibfnamefont
  {M.}~\bibnamefont {Zaidi}},\ }\href@noop {} {\bibfield  {journal} {\bibinfo
  {journal} {J. Alloys Compd.}\ }\textbf {\bibinfo {volume} {580}},\ \bibinfo
  {pages} {558} (\bibinfo {year} {2013})}\BibitemShut {NoStop}%
\bibitem [{\citenamefont {Berger}\ \emph {et~al.}(2002)\citenamefont {Berger},
  \citenamefont {Campillo}, \citenamefont {Vivas}, \citenamefont {Pearson},\
  and\ \citenamefont {Bader}}]{J.Appl.Phys.A.Berger}%
  \BibitemOpen
  \bibfield  {author} {\bibinfo {author} {\bibfnamefont {A.}~\bibnamefont
  {Berger}}, \bibinfo {author} {\bibfnamefont {G.}~\bibnamefont {Campillo}},
  \bibinfo {author} {\bibfnamefont {P.}~\bibnamefont {Vivas}}, \bibinfo
  {author} {\bibfnamefont {J.~E.}\ \bibnamefont {Pearson}}, \ and\ \bibinfo
  {author} {\bibfnamefont {S.~D.}\ \bibnamefont {Bader}},\ }\href@noop {}
  {\bibfield  {journal} {\bibinfo  {journal} {J. Appl. Phys.}\ }\textbf
  {\bibinfo {volume} {91}},\ \bibinfo {pages} {8393} (\bibinfo {year}
  {2002})}\BibitemShut {NoStop}%
\bibitem [{\citenamefont {Sahana}\ \emph {et~al.}(2003)\citenamefont {Sahana},
  \citenamefont {R\"ossler}, \citenamefont {Ghosh}, \citenamefont {Elizabeth},
  \citenamefont {Bhat}, \citenamefont {D\"orr}, \citenamefont {Eckert},
  \citenamefont {Wolf},\ and\ \citenamefont {M\"uller}}]{PhysRevB.68.144408}%
  \BibitemOpen
  \bibfield  {author} {\bibinfo {author} {\bibfnamefont {M.}~\bibnamefont
  {Sahana}}, \bibinfo {author} {\bibfnamefont {U.~K.}\ \bibnamefont
  {R\"ossler}}, \bibinfo {author} {\bibfnamefont {N.}~\bibnamefont {Ghosh}},
  \bibinfo {author} {\bibfnamefont {S.}~\bibnamefont {Elizabeth}}, \bibinfo
  {author} {\bibfnamefont {H.~L.}\ \bibnamefont {Bhat}}, \bibinfo {author}
  {\bibfnamefont {K.}~\bibnamefont {D\"orr}}, \bibinfo {author} {\bibfnamefont
  {D.}~\bibnamefont {Eckert}}, \bibinfo {author} {\bibfnamefont
  {M.}~\bibnamefont {Wolf}}, \ and\ \bibinfo {author} {\bibfnamefont {K.-H.}\
  \bibnamefont {M\"uller}},\ }\href@noop {} {\bibfield  {journal} {\bibinfo
  {journal} {Phys. Rev. B}\ }\textbf {\bibinfo {volume} {68}},\ \bibinfo
  {pages} {144408} (\bibinfo {year} {2003})}\BibitemShut {NoStop}%
\bibitem [{\citenamefont {Ben~Jemaa}\ \emph {et~al.}(2014)\citenamefont
  {Ben~Jemaa}, \citenamefont {Mahmood}, \citenamefont {Ellouze}, \citenamefont
  {Hlil},\ and\ \citenamefont {Halouani}}]{BenJemaa}%
  \BibitemOpen
  \bibfield  {author} {\bibinfo {author} {\bibfnamefont {F.}~\bibnamefont
  {Ben~Jemaa}}, \bibinfo {author} {\bibfnamefont {S.~H.}\ \bibnamefont
  {Mahmood}}, \bibinfo {author} {\bibfnamefont {M.}~\bibnamefont {Ellouze}},
  \bibinfo {author} {\bibfnamefont {E.~K.}\ \bibnamefont {Hlil}}, \ and\
  \bibinfo {author} {\bibfnamefont {F.}~\bibnamefont {Halouani}},\ }\href@noop
  {} {\bibfield  {journal} {\bibinfo  {journal} {J. Mater. Sci.}\ }\textbf
  {\bibinfo {volume} {49}},\ \bibinfo {pages} {6883} (\bibinfo {year}
  {2014})}\BibitemShut {NoStop}%
\bibitem [{\citenamefont {R\"o\ss{}ler}\ \emph {et~al.}(2004)\citenamefont
  {R\"o\ss{}ler}, \citenamefont {R\"o\ss{}ler}, \citenamefont {Nenkov},
  \citenamefont {Eckert}, \citenamefont {Yusuf}, \citenamefont {D\"orr},\ and\
  \citenamefont {M\"uller}}]{PhysRevB.70.104417}%
  \BibitemOpen
  \bibfield  {author} {\bibinfo {author} {\bibfnamefont {S.}~\bibnamefont
  {R\"o\ss{}ler}}, \bibinfo {author} {\bibfnamefont {U.~K.}\ \bibnamefont
  {R\"o\ss{}ler}}, \bibinfo {author} {\bibfnamefont {K.}~\bibnamefont
  {Nenkov}}, \bibinfo {author} {\bibfnamefont {D.}~\bibnamefont {Eckert}},
  \bibinfo {author} {\bibfnamefont {S.~M.}\ \bibnamefont {Yusuf}}, \bibinfo
  {author} {\bibfnamefont {K.}~\bibnamefont {D\"orr}}, \ and\ \bibinfo {author}
  {\bibfnamefont {K.-H.}\ \bibnamefont {M\"uller}},\ }\href@noop {} {\bibfield
  {journal} {\bibinfo  {journal} {Phys. Rev. B}\ }\textbf {\bibinfo {volume}
  {70}},\ \bibinfo {pages} {104417} (\bibinfo {year} {2004})}\BibitemShut
  {NoStop}%
\bibitem [{\citenamefont {Pramanik}\ and\ \citenamefont
  {Banerjee}(2009)}]{PhysRevB.79.214426}%
  \BibitemOpen
  \bibfield  {author} {\bibinfo {author} {\bibfnamefont {A.~K.}\ \bibnamefont
  {Pramanik}}\ and\ \bibinfo {author} {\bibfnamefont {A.}~\bibnamefont
  {Banerjee}},\ }\href@noop {} {\bibfield  {journal} {\bibinfo  {journal}
  {Phys. Rev. B}\ }\textbf {\bibinfo {volume} {79}},\ \bibinfo {pages} {214426}
  (\bibinfo {year} {2009})}\BibitemShut {NoStop}%
\bibitem [{\citenamefont {\mbox{Vasiliu-Doloca}}\ \emph
  {et~al.}(2007)\citenamefont {\mbox{Vasiliu-Doloca}}, \citenamefont {Lynn},
  \citenamefont {Mukovskii}, \citenamefont {Arsenov},\ and\ \citenamefont
  {Shulyatev}}]{J.Appl.Phys.Vasiliu-Doloca}%
  \BibitemOpen
  \bibfield  {author} {\bibinfo {author} {\bibfnamefont {L.}~\bibnamefont
  {\mbox{Vasiliu-Doloca}}}, \bibinfo {author} {\bibfnamefont {J.~W.}\
  \bibnamefont {Lynn}}, \bibinfo {author} {\bibfnamefont {Y.~M.}\ \bibnamefont
  {Mukovskii}}, \bibinfo {author} {\bibfnamefont {A.~A.}\ \bibnamefont
  {Arsenov}}, \ and\ \bibinfo {author} {\bibfnamefont {D.~A.}\ \bibnamefont
  {Shulyatev}},\ }\href@noop {} {\bibfield  {journal} {\bibinfo  {journal} {J.
  Appl. Phys.}\ }\textbf {\bibinfo {volume} {11}},\ \bibinfo {pages} {7342}
  (\bibinfo {year} {2007})}\BibitemShut {NoStop}%
\bibitem [{\citenamefont {Yu}\ \emph {et~al.}(2018)\citenamefont {Yu},
  \citenamefont {Sun}, \citenamefont {Fan}, \citenamefont {Lan}, \citenamefont
  {Zhang}, \citenamefont {Zhu}, \citenamefont {Han}, \citenamefont {Zhang},
  \citenamefont {Ling},\ and\ \citenamefont {Yang}}]{YU2018393}%
  \BibitemOpen
  \bibfield  {author} {\bibinfo {author} {\bibfnamefont {B.}~\bibnamefont
  {Yu}}, \bibinfo {author} {\bibfnamefont {W.}~\bibnamefont {Sun}}, \bibinfo
  {author} {\bibfnamefont {J.}~\bibnamefont {Fan}}, \bibinfo {author}
  {\bibfnamefont {X.}~\bibnamefont {Lan}}, \bibinfo {author} {\bibfnamefont
  {W.}~\bibnamefont {Zhang}}, \bibinfo {author} {\bibfnamefont
  {Y.}~\bibnamefont {Zhu}}, \bibinfo {author} {\bibfnamefont {H.}~\bibnamefont
  {Han}}, \bibinfo {author} {\bibfnamefont {L.}~\bibnamefont {Zhang}}, \bibinfo
  {author} {\bibfnamefont {L.}~\bibnamefont {Ling}}, \ and\ \bibinfo {author}
  {\bibfnamefont {H.}~\bibnamefont {Yang}},\ }\href@noop {} {\bibfield
  {journal} {\bibinfo  {journal} {Mater. Res. Bull.}\ }\textbf {\bibinfo
  {volume} {99}},\ \bibinfo {pages} {393} (\bibinfo {year} {2018})}\BibitemShut
  {NoStop}%
\bibitem [{\citenamefont {Oleaga}\ \emph {et~al.}(2015)\citenamefont {Oleaga},
  \citenamefont {Salazar}, \citenamefont {Ciomaga~Hatnean},\ and\ \citenamefont
  {Balakrishnan}}]{PhysRevB.92.024409}%
  \BibitemOpen
  \bibfield  {author} {\bibinfo {author} {\bibfnamefont {A.}~\bibnamefont
  {Oleaga}}, \bibinfo {author} {\bibfnamefont {A.}~\bibnamefont {Salazar}},
  \bibinfo {author} {\bibfnamefont {M.}~\bibnamefont {Ciomaga~Hatnean}}, \ and\
  \bibinfo {author} {\bibfnamefont {G.}~\bibnamefont {Balakrishnan}},\
  }\href@noop {} {\bibfield  {journal} {\bibinfo  {journal} {Phys. Rev. B}\
  }\textbf {\bibinfo {volume} {92}},\ \bibinfo {pages} {024409} (\bibinfo
  {year} {2015})}\BibitemShut {NoStop}%
\bibitem [{\citenamefont {Zhang}\ \emph {et~al.}(2013)\citenamefont {Zhang},
  \citenamefont {Lampen}, \citenamefont {Phan}, \citenamefont {Yu},
  \citenamefont {Thanh}, \citenamefont {Dan}, \citenamefont {Lam},
  \citenamefont {Srikanth},\ and\ \citenamefont {Phan}}]{ZHANG2013146}%
  \BibitemOpen
  \bibfield  {author} {\bibinfo {author} {\bibfnamefont {P.}~\bibnamefont
  {Zhang}}, \bibinfo {author} {\bibfnamefont {P.}~\bibnamefont {Lampen}},
  \bibinfo {author} {\bibfnamefont {T.}~\bibnamefont {Phan}}, \bibinfo {author}
  {\bibfnamefont {S.}~\bibnamefont {Yu}}, \bibinfo {author} {\bibfnamefont
  {T.}~\bibnamefont {Thanh}}, \bibinfo {author} {\bibfnamefont
  {N.}~\bibnamefont {Dan}}, \bibinfo {author} {\bibfnamefont {V.}~\bibnamefont
  {Lam}}, \bibinfo {author} {\bibfnamefont {H.}~\bibnamefont {Srikanth}}, \
  and\ \bibinfo {author} {\bibfnamefont {M.}~\bibnamefont {Phan}},\ }\href@noop
  {} {\bibfield  {journal} {\bibinfo  {journal} {J. Magn. Magn. Mater.}\
  }\textbf {\bibinfo {volume} {348}},\ \bibinfo {pages} {146} (\bibinfo {year}
  {2013})}\BibitemShut {NoStop}%
\bibitem [{\citenamefont {Phan}\ \emph
  {et~al.}(2014{\natexlab{a}})\citenamefont {Phan}, \citenamefont {Tran},
  \citenamefont {Thanh}, \citenamefont {Yen}, \citenamefont {Thanh},\ and\
  \citenamefont {Yu}}]{PHAN201440}%
  \BibitemOpen
  \bibfield  {author} {\bibinfo {author} {\bibfnamefont {T.-L.}\ \bibnamefont
  {Phan}}, \bibinfo {author} {\bibfnamefont {Q.}~\bibnamefont {Tran}}, \bibinfo
  {author} {\bibfnamefont {P.}~\bibnamefont {Thanh}}, \bibinfo {author}
  {\bibfnamefont {P.}~\bibnamefont {Yen}}, \bibinfo {author} {\bibfnamefont
  {T.}~\bibnamefont {Thanh}}, \ and\ \bibinfo {author} {\bibfnamefont
  {S.}~\bibnamefont {Yu}},\ }\href@noop {} {\bibfield  {journal} {\bibinfo
  {journal} {Solid State Commun.}\ }\textbf {\bibinfo {volume} {184}},\
  \bibinfo {pages} {40} (\bibinfo {year} {2014}{\natexlab{a}})}\BibitemShut
  {NoStop}%
\bibitem [{\citenamefont {Jeddi}\ \emph {et~al.}(2021)\citenamefont {Jeddi},
  \citenamefont {Massoudi}, \citenamefont {Gharsallah}, \citenamefont {Ahmed},
  \citenamefont {Dhahri},\ and\ \citenamefont {Hlil}}]{RSCAdvJeddi}%
  \BibitemOpen
  \bibfield  {author} {\bibinfo {author} {\bibfnamefont {M.}~\bibnamefont
  {Jeddi}}, \bibinfo {author} {\bibfnamefont {J.}~\bibnamefont {Massoudi}},
  \bibinfo {author} {\bibfnamefont {H.}~\bibnamefont {Gharsallah}}, \bibinfo
  {author} {\bibfnamefont {S.~I.}\ \bibnamefont {Ahmed}}, \bibinfo {author}
  {\bibfnamefont {E.}~\bibnamefont {Dhahri}}, \ and\ \bibinfo {author}
  {\bibfnamefont {E.~K.}\ \bibnamefont {Hlil}},\ }\href@noop {} {\bibfield
  {journal} {\bibinfo  {journal} {RSC Adv.}\ }\textbf {\bibinfo {volume}
  {11}},\ \bibinfo {pages} {7238} (\bibinfo {year} {2021})}\BibitemShut
  {NoStop}%
\bibitem [{\citenamefont {Hcini}\ \emph {et~al.}(2015)\citenamefont {Hcini},
  \citenamefont {Boudard}, \citenamefont {Zemni},\ and\ \citenamefont
  {Oumezzine}}]{HCINI20152042}%
  \BibitemOpen
  \bibfield  {author} {\bibinfo {author} {\bibfnamefont {S.}~\bibnamefont
  {Hcini}}, \bibinfo {author} {\bibfnamefont {M.}~\bibnamefont {Boudard}},
  \bibinfo {author} {\bibfnamefont {S.}~\bibnamefont {Zemni}}, \ and\ \bibinfo
  {author} {\bibfnamefont {M.}~\bibnamefont {Oumezzine}},\ }\href@noop {}
  {\bibfield  {journal} {\bibinfo  {journal} {Ceram. Int.}\ }\textbf {\bibinfo
  {volume} {41}},\ \bibinfo {pages} {2042} (\bibinfo {year}
  {2015})}\BibitemShut {NoStop}%
\bibitem [{\citenamefont {Baazaoui}\ \emph {et~al.}(2014)\citenamefont
  {Baazaoui}, \citenamefont {Oumezzine},\ and\ \citenamefont
  {Cheikhrouhou-Koubaa}}]{Phys.SolidStateBaazaoui}%
  \BibitemOpen
  \bibfield  {author} {\bibinfo {author} {\bibfnamefont {M.}~\bibnamefont
  {Baazaoui}}, \bibinfo {author} {\bibfnamefont {M.}~\bibnamefont {Oumezzine}},
  \ and\ \bibinfo {author} {\bibfnamefont {W.}~\bibnamefont
  {Cheikhrouhou-Koubaa}},\ }\href@noop {} {\bibfield  {journal} {\bibinfo
  {journal} {Phys. Solid State}\ }\textbf {\bibinfo {volume} {62}},\ \bibinfo
  {pages} {278} (\bibinfo {year} {2014})}\BibitemShut {NoStop}%
\bibitem [{\citenamefont {Wilson}\ and\ \citenamefont
  {Fisher}(1972)}]{PhysRevLett.28.240}%
  \BibitemOpen
  \bibfield  {author} {\bibinfo {author} {\bibfnamefont {K.~G.}\ \bibnamefont
  {Wilson}}\ and\ \bibinfo {author} {\bibfnamefont {M.~E.}\ \bibnamefont
  {Fisher}},\ }\href@noop {} {\bibfield  {journal} {\bibinfo  {journal} {Phys.
  Rev. Lett.}\ }\textbf {\bibinfo {volume} {28}},\ \bibinfo {pages} {240}
  (\bibinfo {year} {1972})}\BibitemShut {NoStop}%
\bibitem [{\citenamefont {Kaniadakis}\ \emph {et~al.}(2005)\citenamefont
  {Kaniadakis}, \citenamefont {Lissia},\ and\ \citenamefont
  {Scarfone}}]{PhysRevE.71.046128}%
  \BibitemOpen
  \bibfield  {author} {\bibinfo {author} {\bibfnamefont {G.}~\bibnamefont
  {Kaniadakis}}, \bibinfo {author} {\bibfnamefont {M.}~\bibnamefont {Lissia}},
  \ and\ \bibinfo {author} {\bibfnamefont {A.~M.}\ \bibnamefont {Scarfone}},\
  }\href@noop {} {\bibfield  {journal} {\bibinfo  {journal} {Phys. Rev. E}\
  }\textbf {\bibinfo {volume} {71}},\ \bibinfo {pages} {046128} (\bibinfo
  {year} {2005})}\BibitemShut {NoStop}%
\bibitem [{\citenamefont {Kaniadakis}(2009)}]{Eur.Phys.J.B70.3}%
  \BibitemOpen
  \bibfield  {author} {\bibinfo {author} {\bibfnamefont {G.}~\bibnamefont
  {Kaniadakis}},\ }\href@noop {} {\bibfield  {journal} {\bibinfo  {journal}
  {Eur. Phys. J. B}\ }\textbf {\bibinfo {volume} {70}},\ \bibinfo {pages} {3}
  (\bibinfo {year} {2009})}\BibitemShut {NoStop}%
\bibitem [{\citenamefont {Wilson}\ and\ \citenamefont
  {Kogut}(1974)}]{Wilson197475}%
  \BibitemOpen
  \bibfield  {author} {\bibinfo {author} {\bibfnamefont {K.~G.}\ \bibnamefont
  {Wilson}}\ and\ \bibinfo {author} {\bibfnamefont {J.}~\bibnamefont {Kogut}},\
  }\href@noop {} {\bibfield  {journal} {\bibinfo  {journal} {Phys. Rep.}\
  }\textbf {\bibinfo {volume} {12}},\ \bibinfo {pages} {75} (\bibinfo {year}
  {1974})}\BibitemShut {NoStop}%
\bibitem [{\citenamefont {de~Alcantara~Bonfim}\ \emph
  {et~al.}(1981)\citenamefont {de~Alcantara~Bonfim}, \citenamefont {Kirkham},\
  and\ \citenamefont {McKane}}]{Bonfirm_1981}%
  \BibitemOpen
  \bibfield  {author} {\bibinfo {author} {\bibfnamefont {O.~F.}\ \bibnamefont
  {de~Alcantara~Bonfim}}, \bibinfo {author} {\bibfnamefont {J.~E.}\
  \bibnamefont {Kirkham}}, \ and\ \bibinfo {author} {\bibfnamefont {A.~J.}\
  \bibnamefont {McKane}},\ }\href@noop {} {\bibfield  {journal} {\bibinfo
  {journal} {J. Phys. A}\ }\textbf {\bibinfo {volume} {14}},\ \bibinfo {pages}
  {2391} (\bibinfo {year} {1981})}\BibitemShut {NoStop}%
\bibitem [{\citenamefont {Zambelli}\ and\ \citenamefont
  {Zanusso}(2017)}]{PhysRevD.95.085001}%
  \BibitemOpen
  \bibfield  {author} {\bibinfo {author} {\bibfnamefont {L.}~\bibnamefont
  {Zambelli}}\ and\ \bibinfo {author} {\bibfnamefont {O.}~\bibnamefont
  {Zanusso}},\ }\href@noop {} {\bibfield  {journal} {\bibinfo  {journal} {Phys.
  Rev. D}\ }\textbf {\bibinfo {volume} {95}},\ \bibinfo {pages} {085001}
  (\bibinfo {year} {2017})}\BibitemShut {NoStop}%
\bibitem [{\citenamefont {Borinsky}\ \emph {et~al.}(2021)\citenamefont
  {Borinsky}, \citenamefont {Gracey}, \citenamefont {Kompaniets},\ and\
  \citenamefont {Schnetz}}]{PhysRevD.103.116024}%
  \BibitemOpen
  \bibfield  {author} {\bibinfo {author} {\bibfnamefont {M.}~\bibnamefont
  {Borinsky}}, \bibinfo {author} {\bibfnamefont {J.~A.}\ \bibnamefont
  {Gracey}}, \bibinfo {author} {\bibfnamefont {M.~V.}\ \bibnamefont
  {Kompaniets}}, \ and\ \bibinfo {author} {\bibfnamefont {O.}~\bibnamefont
  {Schnetz}},\ }\href@noop {} {\bibfield  {journal} {\bibinfo  {journal} {Phys.
  Rev. D}\ }\textbf {\bibinfo {volume} {103}},\ \bibinfo {pages} {116024}
  (\bibinfo {year} {2021})}\BibitemShut {NoStop}%
\bibitem [{\citenamefont {Stephen}\ and\ \citenamefont
  {McCauley}(1973)}]{STEPHEN197389}%
  \BibitemOpen
  \bibfield  {author} {\bibinfo {author} {\bibfnamefont {M.}~\bibnamefont
  {Stephen}}\ and\ \bibinfo {author} {\bibfnamefont {J.}~\bibnamefont
  {McCauley}},\ }\href@noop {} {\bibfield  {journal} {\bibinfo  {journal}
  {Phys. Lett. A}\ }\textbf {\bibinfo {volume} {44}},\ \bibinfo {pages} {89}
  (\bibinfo {year} {1973})}\BibitemShut {NoStop}%
\bibitem [{\citenamefont {Hager}\ and\ \citenamefont
  {Sch\"afer}(1999)}]{PhysRevE.60.2071}%
  \BibitemOpen
  \bibfield  {author} {\bibinfo {author} {\bibfnamefont {J.}~\bibnamefont
  {Hager}}\ and\ \bibinfo {author} {\bibfnamefont {L.}~\bibnamefont
  {Sch\"afer}},\ }\href@noop {} {\bibfield  {journal} {\bibinfo  {journal}
  {Phys. Rev. E}\ }\textbf {\bibinfo {volume} {60}},\ \bibinfo {pages} {2071}
  (\bibinfo {year} {1999})}\BibitemShut {NoStop}%
\bibitem [{\citenamefont {Hager}(2002)}]{Hager_2002}%
  \BibitemOpen
  \bibfield  {author} {\bibinfo {author} {\bibfnamefont {J.~S.}\ \bibnamefont
  {Hager}},\ }\href@noop {} {\bibfield  {journal} {\bibinfo  {journal} {J.
  Phys. A}\ }\textbf {\bibinfo {volume} {35}},\ \bibinfo {pages} {2703}
  (\bibinfo {year} {2002})}\BibitemShut {NoStop}%
\bibitem [{\citenamefont {Brezin}\ \emph {et~al.}(2014)\citenamefont {Brezin},
  \citenamefont {Parisi},\ and\ \citenamefont
  {Ricci-Tersenghi}}]{BrezinEandParisiGandRicci-TersenghiF}%
  \BibitemOpen
  \bibfield  {author} {\bibinfo {author} {\bibfnamefont {E.}~\bibnamefont
  {Brezin}}, \bibinfo {author} {\bibfnamefont {G.}~\bibnamefont {Parisi}}, \
  and\ \bibinfo {author} {\bibfnamefont {F.}~\bibnamefont {Ricci-Tersenghi}},\
  }\href@noop {} {\bibfield  {journal} {\bibinfo  {journal} {J. Stat. Phys.}\
  }\textbf {\bibinfo {volume} {157}},\ \bibinfo {pages} {855} (\bibinfo {year}
  {2014})}\BibitemShut {NoStop}%
\bibitem [{\citenamefont {Lohmann}\ \emph {et~al.}(2017)\citenamefont
  {Lohmann}, \citenamefont {Slade},\ and\ \citenamefont
  {Lallace}}]{LohmannMSladeGLallaceBC}%
  \BibitemOpen
  \bibfield  {author} {\bibinfo {author} {\bibfnamefont {M.}~\bibnamefont
  {Lohmann}}, \bibinfo {author} {\bibfnamefont {G.}~\bibnamefont {Slade}}, \
  and\ \bibinfo {author} {\bibfnamefont {B.~C.}\ \bibnamefont {Lallace}},\
  }\href@noop {} {\bibfield  {journal} {\bibinfo  {journal} {J. Stat. Phys.}\
  }\textbf {\bibinfo {volume} {169}},\ \bibinfo {pages} {1132} (\bibinfo {year}
  {2017})}\BibitemShut {NoStop}%
\bibitem [{\citenamefont {Slade}(2018)}]{SladeG}%
  \BibitemOpen
  \bibfield  {author} {\bibinfo {author} {\bibfnamefont {G.}~\bibnamefont
  {Slade}},\ }\href@noop {} {\bibfield  {journal} {\bibinfo  {journal} {Commun.
  Math. Phys.}\ }\textbf {\bibinfo {volume} {358}},\ \bibinfo {pages} {343}
  (\bibinfo {year} {2018})}\BibitemShut {NoStop}%
\bibitem [{\citenamefont {Fisher}\ \emph {et~al.}(1972)\citenamefont {Fisher},
  \citenamefont {Ma},\ and\ \citenamefont {Nickel}}]{PhysRevLett.29.917}%
  \BibitemOpen
  \bibfield  {author} {\bibinfo {author} {\bibfnamefont {M.~E.}\ \bibnamefont
  {Fisher}}, \bibinfo {author} {\bibfnamefont {S.-k.}\ \bibnamefont {Ma}}, \
  and\ \bibinfo {author} {\bibfnamefont {B.~G.}\ \bibnamefont {Nickel}},\
  }\href@noop {} {\bibfield  {journal} {\bibinfo  {journal} {Phys. Rev. Lett.}\
  }\textbf {\bibinfo {volume} {29}},\ \bibinfo {pages} {917} (\bibinfo {year}
  {1972})}\BibitemShut {NoStop}%
\bibitem [{\citenamefont {Gross}\ and\ \citenamefont
  {Neveu}(1974)}]{PhysRevD.10.3235}%
  \BibitemOpen
  \bibfield  {author} {\bibinfo {author} {\bibfnamefont {D.~J.}\ \bibnamefont
  {Gross}}\ and\ \bibinfo {author} {\bibfnamefont {A.}~\bibnamefont {Neveu}},\
  }\href@noop {} {\bibfield  {journal} {\bibinfo  {journal} {Phys. Rev. D}\
  }\textbf {\bibinfo {volume} {10}},\ \bibinfo {pages} {3235} (\bibinfo {year}
  {1974})}\BibitemShut {NoStop}%
\bibitem [{\citenamefont {Gracey}\ \emph {et~al.}(2016)\citenamefont {Gracey},
  \citenamefont {Luthe},\ and\ \citenamefont
  {Schr\"oder}}]{PhysRevD.94.125028}%
  \BibitemOpen
  \bibfield  {author} {\bibinfo {author} {\bibfnamefont {J.~A.}\ \bibnamefont
  {Gracey}}, \bibinfo {author} {\bibfnamefont {T.}~\bibnamefont {Luthe}}, \
  and\ \bibinfo {author} {\bibfnamefont {Y.}~\bibnamefont {Schr\"oder}},\
  }\href@noop {} {\bibfield  {journal} {\bibinfo  {journal} {Phys. Rev. D}\
  }\textbf {\bibinfo {volume} {94}},\ \bibinfo {pages} {125028} (\bibinfo
  {year} {2016})}\BibitemShut {NoStop}%
\bibitem [{\citenamefont {Br\'ezin}\ and\ \citenamefont
  {Zinn-Justin}(1976)}]{PhysRevB.13.251}%
  \BibitemOpen
  \bibfield  {author} {\bibinfo {author} {\bibfnamefont {E.}~\bibnamefont
  {Br\'ezin}}\ and\ \bibinfo {author} {\bibfnamefont {J.}~\bibnamefont
  {Zinn-Justin}},\ }\href@noop {} {\bibfield  {journal} {\bibinfo  {journal}
  {Phys. Rev. B}\ }\textbf {\bibinfo {volume} {13}},\ \bibinfo {pages} {251}
  (\bibinfo {year} {1976})}\BibitemShut {NoStop}%
\bibitem [{\citenamefont {Berlin}\ and\ \citenamefont
  {Kac}(1952)}]{PhysRev.86.821}%
  \BibitemOpen
  \bibfield  {author} {\bibinfo {author} {\bibfnamefont {T.~H.}\ \bibnamefont
  {Berlin}}\ and\ \bibinfo {author} {\bibfnamefont {M.}~\bibnamefont {Kac}},\
  }\href@noop {} {\bibfield  {journal} {\bibinfo  {journal} {Phys. Rev.}\
  }\textbf {\bibinfo {volume} {86}},\ \bibinfo {pages} {821} (\bibinfo {year}
  {1952})}\BibitemShut {NoStop}%
\bibitem [{\citenamefont {Hornreich}\ \emph {et~al.}(1975)\citenamefont
  {Hornreich}, \citenamefont {Luban},\ and\ \citenamefont
  {Shtrikman}}]{PhysRevLett.35.1678}%
  \BibitemOpen
  \bibfield  {author} {\bibinfo {author} {\bibfnamefont {R.~M.}\ \bibnamefont
  {Hornreich}}, \bibinfo {author} {\bibfnamefont {M.}~\bibnamefont {Luban}}, \
  and\ \bibinfo {author} {\bibfnamefont {S.}~\bibnamefont {Shtrikman}},\
  }\href@noop {} {\bibfield  {journal} {\bibinfo  {journal} {Phys. Rev. Lett.}\
  }\textbf {\bibinfo {volume} {35}},\ \bibinfo {pages} {1678} (\bibinfo {year}
  {1975})}\BibitemShut {NoStop}%
\bibitem [{\citenamefont {Leite}(2003{\natexlab{a}})}]{PhysRevB.67.104415}%
  \BibitemOpen
  \bibfield  {author} {\bibinfo {author} {\bibfnamefont {M.~M.}\ \bibnamefont
  {Leite}},\ }\href@noop {} {\bibfield  {journal} {\bibinfo  {journal} {Phys.
  Rev. B}\ }\textbf {\bibinfo {volume} {67}},\ \bibinfo {pages} {104415}
  (\bibinfo {year} {2003}{\natexlab{a}})}\BibitemShut {NoStop}%
\bibitem [{\citenamefont {Leite}(2005)}]{PhysRevB.72.224432}%
  \BibitemOpen
  \bibfield  {author} {\bibinfo {author} {\bibfnamefont {M.~M.}\ \bibnamefont
  {Leite}},\ }\href@noop {} {\bibfield  {journal} {\bibinfo  {journal} {Phys.
  Rev. B}\ }\textbf {\bibinfo {volume} {72}},\ \bibinfo {pages} {224432}
  (\bibinfo {year} {2005})}\BibitemShut {NoStop}%
\bibitem [{\citenamefont {de~Albuquerque}\ and\ \citenamefont
  {Leite}(2001)}]{Albuquerque_2001}%
  \BibitemOpen
  \bibfield  {author} {\bibinfo {author} {\bibfnamefont {L.~C.}\ \bibnamefont
  {de~Albuquerque}}\ and\ \bibinfo {author} {\bibfnamefont {M.~M.}\
  \bibnamefont {Leite}},\ }\href@noop {} {\bibfield  {journal} {\bibinfo
  {journal} {J. Phys. A}\ }\textbf {\bibinfo {volume} {34}},\ \bibinfo {pages}
  {L327} (\bibinfo {year} {2001})}\BibitemShut {NoStop}%
\bibitem [{\citenamefont {Leite}(2004)}]{LEITE2004281}%
  \BibitemOpen
  \bibfield  {author} {\bibinfo {author} {\bibfnamefont {M.~M.}\ \bibnamefont
  {Leite}},\ }\href@noop {} {\bibfield  {journal} {\bibinfo  {journal} {Phys.
  Lett. A}\ }\textbf {\bibinfo {volume} {326}},\ \bibinfo {pages} {281}
  (\bibinfo {year} {2004})}\BibitemShut {NoStop}%
\bibitem [{\citenamefont {Leite}(2000)}]{PhysRevB.61.14691}%
  \BibitemOpen
  \bibfield  {author} {\bibinfo {author} {\bibfnamefont {M.~M.}\ \bibnamefont
  {Leite}},\ }\href@noop {} {\bibfield  {journal} {\bibinfo  {journal} {Phys.
  Rev. B}\ }\textbf {\bibinfo {volume} {61}},\ \bibinfo {pages} {14691}
  (\bibinfo {year} {2000})}\BibitemShut {NoStop}%
\bibitem [{\citenamefont {Leite}(2003{\natexlab{b}})}]{PhysRevB.68.052408}%
  \BibitemOpen
  \bibfield  {author} {\bibinfo {author} {\bibfnamefont {M.~M.}\ \bibnamefont
  {Leite}},\ }\href@noop {} {\bibfield  {journal} {\bibinfo  {journal} {Phys.
  Rev. B}\ }\textbf {\bibinfo {volume} {68}},\ \bibinfo {pages} {052408}
  (\bibinfo {year} {2003}{\natexlab{b}})}\BibitemShut {NoStop}%
\bibitem [{\citenamefont {Farias}\ and\ \citenamefont {Leite}(2012)}]{FARIAS}%
  \BibitemOpen
  \bibfield  {author} {\bibinfo {author} {\bibfnamefont {C.~F.}\ \bibnamefont
  {Farias}}\ and\ \bibinfo {author} {\bibfnamefont {M.~M.}\ \bibnamefont
  {Leite}},\ }\href@noop {} {\bibfield  {journal} {\bibinfo  {journal} {J.
  Stat. Phys.}\ }\textbf {\bibinfo {volume} {148}},\ \bibinfo {pages} {972}
  (\bibinfo {year} {2012})}\BibitemShut {NoStop}%
\bibitem [{\citenamefont {da~Silva~Jr.}\ and\ \citenamefont
  {Leite}(2012)}]{Borba}%
  \BibitemOpen
  \bibfield  {author} {\bibinfo {author} {\bibfnamefont {J.~B.}\ \bibnamefont
  {da~Silva~Jr.}}\ and\ \bibinfo {author} {\bibfnamefont {M.~M.}\ \bibnamefont
  {Leite}},\ }\href@noop {} {\bibfield  {journal} {\bibinfo  {journal} {J.
  Math. Phys.}\ }\textbf {\bibinfo {volume} {53}},\ \bibinfo {pages} {043303}
  (\bibinfo {year} {2012})}\BibitemShut {NoStop}%
\bibitem [{\citenamefont {Santos}\ and\ \citenamefont
  {Leite}(2014)}]{Santos_2014}%
  \BibitemOpen
  \bibfield  {author} {\bibinfo {author} {\bibfnamefont {M.~V.~S.}\
  \bibnamefont {Santos}}\ and\ \bibinfo {author} {\bibfnamefont {M.~M.}\
  \bibnamefont {Leite}},\ }\href@noop {} {\bibfield  {journal} {\bibinfo
  {journal} {J. Phys. Conf. Ser.}\ }\textbf {\bibinfo {volume} {490}},\
  \bibinfo {pages} {012232} (\bibinfo {year} {2014})}\BibitemShut {NoStop}%
\bibitem [{\citenamefont {de~Sena}\ and\ \citenamefont
  {Leite}(2015)}]{deSena_2015}%
  \BibitemOpen
  \bibfield  {author} {\bibinfo {author} {\bibfnamefont {M.~I.}\ \bibnamefont
  {de~Sena}}\ and\ \bibinfo {author} {\bibfnamefont {M.~M.}\ \bibnamefont
  {Leite}},\ }\href@noop {} {\bibfield  {journal} {\bibinfo  {journal} {J.
  Phys. Conf. Ser.}\ }\textbf {\bibinfo {volume} {574}},\ \bibinfo {pages}
  {012170} (\bibinfo {year} {2015})}\BibitemShut {NoStop}%
\bibitem [{\citenamefont {Santos}\ \emph
  {et~al.}(2019{\natexlab{a}})\citenamefont {Santos}, \citenamefont
  {da~Silva~Jr},\ and\ \citenamefont {Leite}}]{Santos_2019}%
  \BibitemOpen
  \bibfield  {author} {\bibinfo {author} {\bibfnamefont {M.~V.~S.}\
  \bibnamefont {Santos}}, \bibinfo {author} {\bibfnamefont {J.~B.}\
  \bibnamefont {da~Silva~Jr}}, \ and\ \bibinfo {author} {\bibfnamefont {M.~M.}\
  \bibnamefont {Leite}},\ }\href@noop {} {\bibfield  {journal} {\bibinfo
  {journal} {Eur. Phys. J. Plus}\ }\textbf {\bibinfo {volume} {134}},\ \bibinfo
  {pages} {372} (\bibinfo {year} {2019}{\natexlab{a}})}\BibitemShut {NoStop}%
\bibitem [{\citenamefont {Santos}\ \emph
  {et~al.}(2019{\natexlab{b}})\citenamefont {Santos}, \citenamefont
  {da~Silva~Jr},\ and\ \citenamefont {Leite}}]{Santos_20192}%
  \BibitemOpen
  \bibfield  {author} {\bibinfo {author} {\bibfnamefont {M.~V.~S.}\
  \bibnamefont {Santos}}, \bibinfo {author} {\bibfnamefont {J.~B.}\
  \bibnamefont {da~Silva~Jr}}, \ and\ \bibinfo {author} {\bibfnamefont {M.~M.}\
  \bibnamefont {Leite}},\ }\href@noop {} {\bibfield  {journal} {\bibinfo
  {journal} {Eur. Phys. J. Plus}\ }\textbf {\bibinfo {volume} {134}},\ \bibinfo
  {pages} {4} (\bibinfo {year} {2019}{\natexlab{b}})}\BibitemShut {NoStop}%
\bibitem [{\citenamefont {Leite}(2022)}]{Leite_2022}%
  \BibitemOpen
  \bibfield  {author} {\bibinfo {author} {\bibfnamefont {M.~M.}\ \bibnamefont
  {Leite}},\ }\href@noop {} {\bibfield  {journal} {\bibinfo  {journal} {EPL}\
  }\textbf {\bibinfo {volume} {137}},\ \bibinfo {pages} {34001} (\bibinfo
  {year} {2022})}\BibitemShut {NoStop}%
\bibitem [{\citenamefont {Theumann}\ and\ \citenamefont
  {Gusm\~{a}o}(1985)}]{PhysRevB.31.379}%
  \BibitemOpen
  \bibfield  {author} {\bibinfo {author} {\bibfnamefont {W.~K.}\ \bibnamefont
  {Theumann}}\ and\ \bibinfo {author} {\bibfnamefont {M.~A.}\ \bibnamefont
  {Gusm\~{a}o}},\ }\href@noop {} {\bibfield  {journal} {\bibinfo  {journal}
  {Phys. Rev. B}\ }\textbf {\bibinfo {volume} {31}},\ \bibinfo {pages} {379}
  (\bibinfo {year} {1985})}\BibitemShut {NoStop}%
\bibitem [{\citenamefont {Itzykson}\ and\ \citenamefont
  {Drouffe}(1989)}]{C.ItzyksonJ.M.Drouffe}%
  \BibitemOpen
  \bibfield  {author} {\bibinfo {author} {\bibfnamefont {C.}~\bibnamefont
  {Itzykson}}\ and\ \bibinfo {author} {\bibfnamefont {J.~M.}\ \bibnamefont
  {Drouffe}},\ }\href@noop {} {\emph {\bibinfo {title} {Statistical Field
  Theory: Vol. 1}}}\ (\bibinfo  {publisher} {Cambridge Monographs on
  Mathematical Physics},\ \bibinfo {year} {1989})\BibitemShut {NoStop}%
\bibitem [{\citenamefont {Zinn-Justin}(1991)}]{ZINNJUSTIN1991105}%
  \BibitemOpen
  \bibfield  {author} {\bibinfo {author} {\bibfnamefont {J.}~\bibnamefont
  {Zinn-Justin}},\ }\href@noop {} {\bibfield  {journal} {\bibinfo  {journal}
  {Nucl. Phys. B}\ }\textbf {\bibinfo {volume} {367}},\ \bibinfo {pages} {105}
  (\bibinfo {year} {1991})}\BibitemShut {NoStop}%
\bibitem [{\citenamefont {Zerf}\ \emph {et~al.}(2017)\citenamefont {Zerf},
  \citenamefont {Mihaila}, \citenamefont {Marquard}, \citenamefont {Herbut},\
  and\ \citenamefont {Scherer}}]{PhysRevD.96.096010}%
  \BibitemOpen
  \bibfield  {author} {\bibinfo {author} {\bibfnamefont {N.}~\bibnamefont
  {Zerf}}, \bibinfo {author} {\bibfnamefont {L.~N.}\ \bibnamefont {Mihaila}},
  \bibinfo {author} {\bibfnamefont {P.}~\bibnamefont {Marquard}}, \bibinfo
  {author} {\bibfnamefont {I.~F.}\ \bibnamefont {Herbut}}, \ and\ \bibinfo
  {author} {\bibfnamefont {M.~M.}\ \bibnamefont {Scherer}},\ }\href@noop {}
  {\bibfield  {journal} {\bibinfo  {journal} {Phys. Rev. D}\ }\textbf {\bibinfo
  {volume} {96}},\ \bibinfo {pages} {096010} (\bibinfo {year}
  {2017})}\BibitemShut {NoStop}%
\bibitem [{\citenamefont {\mbox{Hans-Karl} Janssen}\ and\ \citenamefont
  {Täuber}(2005)}]{JANSSEN2005147}%
  \BibitemOpen
  \bibfield  {author} {\bibinfo {author} {\bibnamefont {\mbox{Hans-Karl}
  Janssen}}\ and\ \bibinfo {author} {\bibfnamefont {U.~C.}\ \bibnamefont
  {Täuber}},\ }\href@noop {} {\bibfield  {journal} {\bibinfo  {journal} {Ann.
  Phys.}\ }\textbf {\bibinfo {volume} {315}},\ \bibinfo {pages} {147} (\bibinfo
  {year} {2005})}\BibitemShut {NoStop}%
\bibitem [{\citenamefont {Täuber}\ \emph {et~al.}(2005)\citenamefont
  {Täuber}, \citenamefont {Howard},\ and\ \citenamefont
  {Vollmayr-Lee}}]{Tauber_2005}%
  \BibitemOpen
  \bibfield  {author} {\bibinfo {author} {\bibfnamefont {U.~C.}\ \bibnamefont
  {Täuber}}, \bibinfo {author} {\bibfnamefont {M.}~\bibnamefont {Howard}}, \
  and\ \bibinfo {author} {\bibfnamefont {B.~P.}\ \bibnamefont {Vollmayr-Lee}},\
  }\href@noop {} {\bibfield  {journal} {\bibinfo  {journal} {J. Phys. A Math.
  Theor.}\ }\textbf {\bibinfo {volume} {38}},\ \bibinfo {pages} {R79} (\bibinfo
  {year} {2005})}\BibitemShut {NoStop}%
\bibitem [{\citenamefont {Stanley}(1968)}]{PhysRev.176.718}%
  \BibitemOpen
  \bibfield  {author} {\bibinfo {author} {\bibfnamefont {H.~E.}\ \bibnamefont
  {Stanley}},\ }\href@noop {} {\bibfield  {journal} {\bibinfo  {journal} {Phys.
  Rev.}\ }\textbf {\bibinfo {volume} {176}},\ \bibinfo {pages} {718} (\bibinfo
  {year} {1968})}\BibitemShut {NoStop}%
\bibitem [{\citenamefont {Phan}\ \emph {et~al.}(2010)\citenamefont {Phan},
  \citenamefont {Franco}, \citenamefont {Bingham}, \citenamefont {Srikanth},
  \citenamefont {Hur},\ and\ \citenamefont {Yu}}]{PHAN2010238}%
  \BibitemOpen
  \bibfield  {author} {\bibinfo {author} {\bibfnamefont {M.}~\bibnamefont
  {Phan}}, \bibinfo {author} {\bibfnamefont {V.}~\bibnamefont {Franco}},
  \bibinfo {author} {\bibfnamefont {N.}~\bibnamefont {Bingham}}, \bibinfo
  {author} {\bibfnamefont {H.}~\bibnamefont {Srikanth}}, \bibinfo {author}
  {\bibfnamefont {N.}~\bibnamefont {Hur}}, \ and\ \bibinfo {author}
  {\bibfnamefont {S.}~\bibnamefont {Yu}},\ }\href@noop {} {\bibfield  {journal}
  {\bibinfo  {journal} {J. Alloys Compd.}\ }\textbf {\bibinfo {volume} {508}},\
  \bibinfo {pages} {238} (\bibinfo {year} {2010})}\BibitemShut {NoStop}%
\bibitem [{\citenamefont {Nisha}\ \emph
  {et~al.}(2012{\natexlab{a}})\citenamefont {Nisha}, \citenamefont {{Savitha
  Pillai}}, \citenamefont {Darbandi}, \citenamefont {Varma}, \citenamefont
  {Suresh},\ and\ \citenamefont {Hahn}}]{NISHA201266}%
  \BibitemOpen
  \bibfield  {author} {\bibinfo {author} {\bibfnamefont {P.}~\bibnamefont
  {Nisha}}, \bibinfo {author} {\bibfnamefont {S.}~\bibnamefont {{Savitha
  Pillai}}}, \bibinfo {author} {\bibfnamefont {A.}~\bibnamefont {Darbandi}},
  \bibinfo {author} {\bibfnamefont {M.~R.}\ \bibnamefont {Varma}}, \bibinfo
  {author} {\bibfnamefont {K.}~\bibnamefont {Suresh}}, \ and\ \bibinfo {author}
  {\bibfnamefont {H.}~\bibnamefont {Hahn}},\ }\href@noop {} {\bibfield
  {journal} {\bibinfo  {journal} {Mater. Chem. Phys.}\ }\textbf {\bibinfo
  {volume} {136}},\ \bibinfo {pages} {66} (\bibinfo {year}
  {2012}{\natexlab{a}})}\BibitemShut {NoStop}%
\bibitem [{\citenamefont {Nisha}\ \emph
  {et~al.}(2012{\natexlab{b}})\citenamefont {Nisha}, \citenamefont {{Savitha
  Pillai}}, \citenamefont {Varma},\ and\ \citenamefont {Suresh}}]{NISHA201240}%
  \BibitemOpen
  \bibfield  {author} {\bibinfo {author} {\bibfnamefont {P.}~\bibnamefont
  {Nisha}}, \bibinfo {author} {\bibfnamefont {S.}~\bibnamefont {{Savitha
  Pillai}}}, \bibinfo {author} {\bibfnamefont {M.~R.}\ \bibnamefont {Varma}}, \
  and\ \bibinfo {author} {\bibfnamefont {K.}~\bibnamefont {Suresh}},\
  }\href@noop {} {\bibfield  {journal} {\bibinfo  {journal} {Solid State Sci.}\
  }\textbf {\bibinfo {volume} {14}},\ \bibinfo {pages} {40} (\bibinfo {year}
  {2012}{\natexlab{b}})}\BibitemShut {NoStop}%
\bibitem [{\citenamefont {Bary}(2022)}]{BARY2022112572}%
  \BibitemOpen
  \bibfield  {author} {\bibinfo {author} {\bibfnamefont {G.}~\bibnamefont
  {Bary}},\ }\href@noop {} {\bibfield  {journal} {\bibinfo  {journal} {Chaos,
  Solitons \& Fractals}\ }\textbf {\bibinfo {volume} {164}},\ \bibinfo {pages}
  {112572} (\bibinfo {year} {2022})}\BibitemShut {NoStop}%
\bibitem [{\citenamefont {Arrott}\ and\ \citenamefont
  {Noakes}(1967)}]{PhysRevLett.19.786}%
  \BibitemOpen
  \bibfield  {author} {\bibinfo {author} {\bibfnamefont {A.}~\bibnamefont
  {Arrott}}\ and\ \bibinfo {author} {\bibfnamefont {J.~E.}\ \bibnamefont
  {Noakes}},\ }\href@noop {} {\bibfield  {journal} {\bibinfo  {journal} {Phys.
  Rev. Lett.}\ }\textbf {\bibinfo {volume} {19}},\ \bibinfo {pages} {786}
  (\bibinfo {year} {1967})}\BibitemShut {NoStop}%
\bibitem [{\citenamefont {Kouvel}\ and\ \citenamefont
  {Fisher}(1964)}]{PhysRev.136.A1626}%
  \BibitemOpen
  \bibfield  {author} {\bibinfo {author} {\bibfnamefont {J.~S.}\ \bibnamefont
  {Kouvel}}\ and\ \bibinfo {author} {\bibfnamefont {M.~E.}\ \bibnamefont
  {Fisher}},\ }\href@noop {} {\bibfield  {journal} {\bibinfo  {journal} {Phys.
  Rev.}\ }\textbf {\bibinfo {volume} {136}},\ \bibinfo {pages} {A1626}
  (\bibinfo {year} {1964})}\BibitemShut {NoStop}%
\bibitem [{\citenamefont {Mnefgui}\ \emph {et~al.}(2014)\citenamefont
  {Mnefgui}, \citenamefont {Zaidi}, \citenamefont {Dhahri}, \citenamefont
  {Hlil},\ and\ \citenamefont {Dhahri}}]{MNEFGUI2014193}%
  \BibitemOpen
  \bibfield  {author} {\bibinfo {author} {\bibfnamefont {S.}~\bibnamefont
  {Mnefgui}}, \bibinfo {author} {\bibfnamefont {N.}~\bibnamefont {Zaidi}},
  \bibinfo {author} {\bibfnamefont {A.}~\bibnamefont {Dhahri}}, \bibinfo
  {author} {\bibfnamefont {E.}~\bibnamefont {Hlil}}, \ and\ \bibinfo {author}
  {\bibfnamefont {J.}~\bibnamefont {Dhahri}},\ }\href@noop {} {\bibfield
  {journal} {\bibinfo  {journal} {J. Solid State Chem.}\ }\textbf {\bibinfo
  {volume} {215}},\ \bibinfo {pages} {193} (\bibinfo {year}
  {2014})}\BibitemShut {NoStop}%
\bibitem [{\citenamefont {Ginting}\ \emph {et~al.}(2013)\citenamefont
  {Ginting}, \citenamefont {Nanto}, \citenamefont {Zhang}, \citenamefont {Yu},\
  and\ \citenamefont {Phan}}]{GINTING201317}%
  \BibitemOpen
  \bibfield  {author} {\bibinfo {author} {\bibfnamefont {D.}~\bibnamefont
  {Ginting}}, \bibinfo {author} {\bibfnamefont {D.}~\bibnamefont {Nanto}},
  \bibinfo {author} {\bibfnamefont {Y.}~\bibnamefont {Zhang}}, \bibinfo
  {author} {\bibfnamefont {S.}~\bibnamefont {Yu}}, \ and\ \bibinfo {author}
  {\bibfnamefont {T.}~\bibnamefont {Phan}},\ }\href@noop {} {\bibfield
  {journal} {\bibinfo  {journal} {Phys. B: Condens. Matter}\ }\textbf {\bibinfo
  {volume} {412}},\ \bibinfo {pages} {17} (\bibinfo {year} {2013})}\BibitemShut
  {NoStop}%
\bibitem [{\citenamefont {Phan}\ \emph
  {et~al.}(2014{\natexlab{b}})\citenamefont {Phan}, \citenamefont {Thanh},\
  and\ \citenamefont {Yu}}]{PHAN2014S247}%
  \BibitemOpen
  \bibfield  {author} {\bibinfo {author} {\bibfnamefont {T.-L.}\ \bibnamefont
  {Phan}}, \bibinfo {author} {\bibfnamefont {T.}~\bibnamefont {Thanh}}, \ and\
  \bibinfo {author} {\bibfnamefont {S.}~\bibnamefont {Yu}},\ }\href@noop {}
  {\bibfield  {journal} {\bibinfo  {journal} {J. Alloys Compd.}\ }\textbf
  {\bibinfo {volume} {615}},\ \bibinfo {pages} {S247} (\bibinfo {year}
  {2014}{\natexlab{b}})}\BibitemShut {NoStop}%
\end{thebibliography}%

\end{document}